\documentclass[onecolumn,showpacs,amsmath,amssymb,12pt]{revtex4}
\usepackage{graphicx}
\usepackage{bm}
\begin{document}
\newcommand{\hs}{\hspace*{0.5cm}}
\newcommand{\vs}{\vspace*{0.5cm}}
\newcommand{\be}{\begin{equation}}
\newcommand{\ee}{\end{equation}}
\newcommand{\bea}{\begin{eqnarray}}
\newcommand{\eea}{\end{eqnarray}}
\newcommand{\ben}{\begin{enumerate}}
\newcommand{\een}{\end{enumerate}}
\newcommand{\bde}{\begin{widetext}}
\newcommand{\ede}{\end{widetext}}
\newcommand{\nn}{\nonumber}
\newcommand{\crn}{\nonumber \\}
\newcommand{\Tr}{\mathrm{Tr}}
\newcommand{\non}{\nonumber}
\newcommand{\noi}{\noindent}
\newcommand{\al}{\alpha}
\newcommand{\la}{\lambda}
\newcommand{\bet}{\beta}
\newcommand{\ga}{\gamma}
\newcommand{\va}{\varphi}
\newcommand{\om}{\omega}
\newcommand{\pa}{\partial}
\newcommand{\+}{\dagger}
\newcommand{\fr}{\frac}
\newcommand{\bc}{\begin{center}}
\newcommand{\ec}{\end{center}}
\newcommand{\Ga}{\Gamma}
\newcommand{\de}{\delta}
\newcommand{\De}{\Delta}
\newcommand{\ep}{\epsilon}
\newcommand{\varep}{\varepsilon}
\newcommand{\ka}{\kappa}
\newcommand{\La}{\Lambda}
\newcommand{\si}{\sigma}
\newcommand{\Si}{\Sigma}
\newcommand{\ta}{\tau}
\newcommand{\up}{\upsilon}
\newcommand{\Up}{\Upsilon}
\newcommand{\ze}{\zeta}
\newcommand{\ps}{\psi}
\newcommand{\Ps}{\Psi}
\newcommand{\ph}{\phi}
\newcommand{\vph}{\varphi}
\newcommand{\Ph}{\Phi}
\newcommand{\Om}{\Omega}
\title{\bf{
Investigation of Dark Matter \\
in Minimal 3-3-1 Models}}
\author{P. V. Dong\footnote{pvdong@iop.vast.ac.vn}}\affiliation{Institute of Physics, Vietnam Academy of Science and Technology,
 10 Dao Tan, Ba Dinh, Hanoi, Vietnam}

\author{C. S. Kim\footnote{Corresponding author, ~~ cskim@yonsei.ac.kr}}\affiliation{Department of Physics and IPAP, Yonsei University, Seoul 120-479, Korea}

\author{D. V. Soa\footnote{dvsoa@assoc.iop.vast.ac.vn}}\affiliation{Department of Physics, Hanoi Metropolitan University,
 98 Duong Quang Ham, Cau Giay, Hanoi, Vietnam}

\author{N. T. Thuy\footnote{ntthuy@iop.vast.ac.vn}}\affiliation{Department of Physics and IPAP, Yonsei University, Seoul 120-479, Korea}

\begin{abstract}
It is shown that the 3-3-1 model with the minimal lepton content can work as two-Higgs-triplet 3-3-1 model while leaving the other scalars as inert particles responsible for dark matter. We study two cases of dark matter corresponding to the doublet and singlet scalar candidates. We figure out the parameter spaces in the WMAP allowed region of the relic density. The indirect and direct searches
for dark matter in both cases are investigated by using micrOMEGAs.
\end{abstract}


 \maketitle

\newpage
\section{\label{intro}Introduction}

Cosmological observations \cite{wmap} suggest that  there must  exist cold dark matter
contained approximately 27$\%$ of all
energy density of the Universe. Dark matter is a mysterious and an interesting
subject in particle physics as well as in astrophysics. In the context of particle
physics, the most popular
dark matter candidates perhaps include the
lightest supersymmetric particle, the lightest KK particle, the lightest $T$-odd particle, the axion, some form of sterile neutrinos, inert scalars, and the others \cite{DM}.

The Standard Model   is very successful in describing experimentally observed phenomena,
but it leaves some unsolved problems,
such as neutrino masses and mixing, matter-antimatter asymmetry, dark matter, dark energy and $etc.$, which guides us to go beyond the Standard Model.
One simple way to go beyond the Standard Model is that
we extend the gauge group $SU(2)_L\otimes U(1)_Y$ to $SU(3)_L\otimes U(1)_X$ \cite{331m,331r}. The class of $SU(3)_C \otimes SU(3)_L \otimes U(1)_X$ (3-3-1) models has many interesting
characteristics since they can explain the number of fermion generations,
the uncharacteristically-heavy top quark~\cite{tquark}, the electric charge quantization \cite{ecq},
the light neutrino masses \cite{neutrino331}, and dark matter \cite{dm331}.

There are two main versions of the 3-3-1 model depending on which type of
particles is located at the bottom of the lepton triplets. The minimal 3-3-1 model \cite{331m} uses ordinary charged leptons
$e_R$, while the version with right-handed neutrinos includes $\nu_R$ \cite{331r}.
There is no dark matter candidate in the original minimal 3-3-1 model, neither in the original 3-3-1 model
with right-handed neutrinos since the new particles in these models are electrically charged or  rapidly
decay. A natural approach \cite{dongdm} is that the stability of dark matter is
based on $W$ parity (similar to $R$ parity in Supersymmetry) by considering
the baryon minus lepton numbers as a local gauge symmetry.
However, this mechanism works only with
the 3-3-1 model with neutral fermions ($N_R$) that possess $L(N_R)=0$ and $B(N_R)=0$. Therefore, the issue of dark matter
for the original 3-3-1 models remains unresolved.

If the $B-L$ charge (even for similar charges that do not commute with $SU(3)_L$) is conserved, the 3-3-1 models are not self-consistent, because the $B-L$ and 3-3-1 symmetries are algebraically non-closed \cite{dongdm,inert331}. Hence, the 3-3-1 models are manifest only if they contain interactions that explicitly violate $B-L$ (which regards $B-L$ as an approximate symmetry). Because the normal Lagrangians of the 3-3-1 models, including the gauge interactions, minimal Yukawa Lagrangian, and minimal scalar potential, conserve $B-L$, the unwanted (abnormal) interactions that violate $B-L$ must present. Such an interaction provides the nonzero, small masses for the neutrinos \cite{DNS}. In this work, we argue that the existence of inert fields can not only make the 3-3-1 model viable but also provide realistic candidates for dark matter. In more detail, one might introduce a $Z_2$ symmetry so that one scalar triplet of the theory is odd, while
all other fields are even under the $Z_2$ symmetry: Odd particles
act as inert fields \cite{idmma}. Therefore, the lightest and neutral inert particle
is stable and can be a dark matter \cite{inert331,DNS}. The inert fields communicate with the normal fields via an interaction that violates $B-L$. This interaction subsequently separates the masses of the inert fields that make the dark matter candidate viable under the direct searches.

The minimal 3-3-1 model originally works with three scalar triplets $\rho=(\rho_1^+,\rho_2^0,\rho_3^{++})$,
$\eta=(\eta_1^0,\eta_2^-,\eta_3^+)$, $\chi=(\chi_1^-, \chi_2^{--},\chi_3^0)$, and either with or without one scalar sextet $S=(S^0_{11},S^-_{12},S^+_{13},S^{--}_{22},S^0_{23},S^{++}_{33})$.
In order to enrich the inert scalar sector responsible for dark matter, one can consider the ``reduced 3-3-1 model"
\cite{r331} by excluding $\eta$ and $S$, or the ``simple 3-3-1 model" \cite{DNS} by excluding $\rho$ and $S$.
Unfortunately, the reduced 3-3-1 model
gives large flavor-changing neutral currents as well as large $\rho$-parameter because the new physics scale is limited by a low Landau pole of around $5$ TeV. The approach with the simple 3-3-1 model seems to be more realistic, except the discrepancy between the FCNC and $\rho$-parameter constraints (however, this has  not really ruled the model out)
\cite{dongsi}.
Additional inert scalars can be a triplet  $\rho$ or sextets ($S$, $\sigma$) or
a replication of $\eta$ or of $\chi$. Among these proposals, the simple 3-3-1 model
with inert scalar sextet $\sigma$ (that has $X=1$ where $X$ is the charge of $U(1)_X$) or with
the replication of $\eta$ or of $\chi$ can provide realistic dark matter candidates. Dark matter candidates for the model with inert
$\sigma$ has already been studied in \cite{DNS}. In this work, we focus on dark matters in the models
with $\eta$ and $\chi$ replications. Let us remind that the dark matter candidates of the model with $\rho$ and the model with $S$ are ruled out by the direct search constraints. Here, in these cases the candidates are degenerate in masses, and the interactions of inert and normal sectors conserve $B-L$ \cite{DNS}.

As a result of $SU(3)_L\otimes U(1)_X$ symmetry, the normal interactions generally produce relevant, new particles in pairs, similarly to superparticles in the Supersymmetry (cf. \cite{dongdm}). Therefore, the 3-3-1 models have been thought to provide dark matter candidates similarly \cite{dm331}. However, the problem is how to suppress or evade the unwanted interactions and vacuums that cause the fast decay of dark matter. The first article in \cite{dm331} discussed a scalar sector of the minimal 3-3-1 model, but the claimed candidate turns out to be the Goldstone boson of $Z'$, which is unstable. Even, the corresponding Higgs field interpreted therein would decay into ordinary particles via its coupling to the Standard Model Higgs bosons, exotic quarks, and gauge bosons. The second and third articles in \cite{dm331} discussed the scalar sector of the 3-3-1 model with right-handed neutrinos, and the candidate was the real or imaginary part of a neutral scalar bilepton. Since the dark matter stability mechanism was not given, there is no reason why the bilepton cannot develop a VEV, and the lepton-number violating (renormalizable) interactions in Yukawa Lagrangian and scalar potential will turn on. Thus, the real part will decay into ordinary particles via the coupling to the Standard Model Higgs bosons, while the real and imaginary parts decay into light quarks due to ordinary and exotic quark mixings. To keep the bilepton stable, the fourth article of \cite{dm331} imposed the lepton number symmetry, which subsequently suppressed all those unwanted interactions and vacuums. However, the problem was to generate the neutrino masses, which finally breaks or violates the symmetry (contradiction to the postulate), and this destabilizes the candidate [e.g., the five-dimensional interactions for neutrino masses mentioned therein will lead to dark matter decays into light neutrinos]. The fifth article of \cite{dm331} introduced another lepton sector, along with a $Z_2$ symmetry or $U(1)_G$ for dark matter stability. But, the $Z_2$ is broken by the Higgs vacuum, while $U(1)_G$ is broken by its nontrivial dynamics \cite{dongdm}. The correct stability mechanism should be a $W$-parity as residual gauge symmetry. However, it works only with a new lepton sector as well as including $B-L$ as a gauge symmetry. To conclude, the dark matter identification and its stability for the typical 3-3-1 models remain unsolved, which have called for our attention. The advantage of inert fields is that the dark mater and neutrino masses can be simultaneously understood.

Our paper is organized as follows: In section \ref{review}, we briefly describe minimal 3-3-1 models that
behave as the simple 3-3-1 model and the versions with $\eta$ and $\chi$ replications.
We also calculate the interactions of the inert particles with the normal matter sector.
In section \ref{darkmatter}, we present the dark matter relic density and experimental searches
for those two models. Finally, we summarize our work in section \ref{conclusion}.

\section{\label{review}Brief description of minimal 3-3-1 models}

\subsection{The simple 3-3-1 model}

The fermions of the simple 3-3-1 model are arranged as \cite{DNS}
\bea \psi_{aL} &\equiv & \left(\begin{array}{c}
               \nu_{aL}  \\ e_{aL} \\ (e_{aR})^c
\end{array}\right) \sim (1,3,0),\crn
Q_{\al L}  &\equiv& \left(\begin{array}{c}
  d_{\al L}\\  -u_{\al L}\\  J_{\al L}
\end{array}\right)\sim (3,3^*,-1/3),\hs Q_{3L} \equiv \left(\begin{array}{c} u_{3L}\\  d_{3L}\\ J_{3L} \end{array}\right)\sim
 \left(3,3,2/3\right), \\ u_{a
R}&\sim&\left(3,1,2/3\right),\hs d_{a R} \sim \left(3,1,-1/3\right),\crn
J_{\al R} &\sim&
\left(3,1,-4/3\right),\hs J_{3R} \sim \left(3,1,5/3\right),\eea where $a=1,2,3$
and $\al= 1,2$ are family indices. The quantum numbers in parentheses are defined
upon the gauge symmetries ($SU(3)_C, SU(3)_L, U(1)_X$), respectively.

The electric charge operator has the form $Q=T_3-\sqrt3 T_8+X$,
where $T_i (i=1,2...,8)$ and $X$ are the charges of $SU(3)_L$
and $U(1)_X$, correspondingly.
The exotic quarks have electric
charges different from the usual ones, $Q(J_\al)=-4/3$ and $Q(J_3)=5/3$.

The model works well with two scalar triplets \cite{DNS} as
 \bea \eta = \left(\begin{array}{c}
\fr 1 {\sqrt2} (u+S_1+i A_1)\\
\eta^-_2\\
\eta^{+}_3\end{array}\right)\sim (1,3,0),\hs \chi = \left(\begin{array}{c}
\chi^-_1\\
\chi^{--}_2\\
\fr 1 {\sqrt2} (\om+S_3+i A_3)\end{array}\right)\sim (1,3,-1).\label{vev2}
\eea

The scalar potential is given by
\be V_{\mathrm{simple}}=\mu^2_1\eta^\dagger \eta + \mu^2_2 \chi^\dagger \chi + \la_1 (\eta^\dagger \eta)^2+\la_2 (\chi^\dagger \chi)^2+\la_3(\eta^\dagger\eta)(\chi^\dagger \chi) +\la_4(\eta^\dagger \chi)(\chi^\dagger\eta),
\label{Vsimple}\ee where $\mu_{1,2}$ have dimension of mass, while $\la_{1,2,3,4}$ are dimensionless.
These parameters satisfy
\be \mu^2_{1,2}<0,\hs \la_{1,2,4}>0,\hs -2\sqrt{\la_1\la_2}<\la_3<\mathrm{Min}\left\{2\la_1\left({\mu_2}/{\mu_1}\right)^2,2\la_2\left({\mu_1}/{\mu_2}\right)^2\right\}.\ee

The model contains four massive scalars with respective masses were obtained in \cite{DNS} as follows
\bea && h \equiv c_\xi S_1-s_\xi S_3,\hs m^2_h\simeq \fr{4\la_1\la_2-\la^2_3}{2\la_2}u^2,\crn
 && H \equiv s_\xi S_1 + c_\xi S_3,\hs m^2_{H}\simeq 2\la_2 \om^2,\\
 && H^{\pm}\equiv c_{\theta}\eta^\pm_3+s_{\theta}\chi^\pm_1,\hs m^2_{H^\pm}\simeq \fr{\la_4}{2} \om^2,\nn \eea
with denotation $c_x=\cos(x),\ s_x=\sin(x),\ t_x=\tan(x)$  for any $x$ angle.
The  mixing angles $\xi$,  $\theta$ are defined as
 \be t_\theta=\fr{u}{\om},\hs t_{2\xi}\simeq \fr{\la_3 u}{\la_2 \om}.\ee

There are eight Goldstone bosons $G_Z\equiv A_1$, $G_{Z'}\equiv A_3$, $G^\pm_{W}\equiv \eta_2^\pm$, $G^{\pm\pm}_{Y}\equiv \chi^{\pm\pm}_2$ and $G^\pm_X\equiv c_\theta \chi^\pm_1-s_\theta \eta_3^\pm$ eaten by  eight massive gauge bosons $Z$, $Z'$, $W^\pm$, $Y^{\pm\pm}$ and $X^\pm$ (see below), correspondingly.
In the limit $u\ll \om$, we have $\xi, \theta \rightarrow 0$, thus
\be h \simeq  S_1, \hs H \simeq  S_3,\hs H^{\pm}\simeq \eta^\pm_3,\hs G^\pm_X\simeq   \chi^\pm_1.  \ee

In the gauge sector, the gauge boson masses arise from the Lagrangian  $\sum_{\Phi = \eta,\chi}(D_\mu\langle \Phi \rangle)^\dagger (D^\mu\langle \Phi \rangle)$, where
the covariant derivative is defined as $D_\mu=\pa_\mu + i g_s t_i G_{i\mu} + i g T_i A_{i\mu} + i g_X X B_\mu$,
in which the gauge coupling constants $g_s,\ g$ and $g_X$ and the gauge bosons $G_{i\mu},\ A_{i\mu}$ and $B_\mu$ are  associated with the 3-3-1 groups, respectively.
The gauge bosons with their masses  are respectively given by \cite{DNS}
\bea &&W^{\pm}\equiv \fr{A_1\mp i A_2}{\sqrt{2}},\hs m^2_W=\fr{g^2}{4}u^2,\crn
&&X^\mp \equiv \fr{A_4 \mp i A_5}{\sqrt{2}},\hs m^2_X=\fr{g^2}{4}(\om^2+u^2),\\
&& Y^{\mp\mp}\equiv \fr{A_6 \mp i A_7}{\sqrt{2}},\hs m^2_Y=\fr{g^2}{4}\om^2, \eea
and for the neutral gauge bosons
\bea &&A = s_W A_{3}+ c_W\left(-\sqrt{3}t_W A_{8}+\sqrt{1-3t^2_W}B\right), \hs m_A=0,\crn
&&Z_1\simeq c_W A_{3 }- s_W\left(-\sqrt{3}t_W A_{8 }+\sqrt{1-3t^2_W}B \right),\hs
m^2_{Z_1}\simeq \fr{g^2}{4c^2_W}u^2,\crn
&& Z_2\simeq\sqrt{1-3t^2_W}A_{8 }+\sqrt{3}t_W B, \hs
m^2_{Z_2} \simeq \fr{g^2c^2_W}{3(1-4s^2_W)}\om^2,\eea
where $s_W=e/g = t/\sqrt{1+4t^2}$, with $t=g_X/g$, is the sine of Weinberg angle \cite{dl}.
The photon field $A_\mu$ is exactly massless. For the gauge bosons $Z_{1 }, Z_{2 }$ we have
taken the limit $u\ll \om$. The $Z_{1 }$ is identified as the Standard Model $Z$. The VEV $u$
is constrained by the mass of $W$, thus $u\simeq 246$ GeV.

The Yukawa Lagrangian is given by
\bea \mathcal{L}_Y&=&h^J_{33}\bar{Q}_{3L}\chi J_{3R}+ h^J_{\al \beta}\bar{Q}_{\al L} \chi^* J_{\beta R}\crn
&&+h^u_{3 a} \bar{Q}_{3L} \eta u_{aR}+ \fr{h^u_{\al a}}{\La} \bar{Q}_{\al L} \eta \chi u_{a R}\crn
&&+ h^d_{\al a} \bar{Q}_{\al L} \eta^* d_{a R} + \fr{h^d_{3 a}}{\La} \bar{Q}_{3L}\eta^*\chi^* d_{a R}  \crn
&&+h^e_{ab} \bar{\psi}^c_{aL} \psi_{bL}\eta + \fr{h'^e_{ab}}{\La^2}(\bar{\psi}^c_{aL}\eta\chi)(\psi_{bL}\chi^*)\crn
&&+\fr{s^\nu_{ab}}{\La} (\bar{\psi}^c_{aL}\eta^*)(\psi_{bL} \eta^*)+\mathrm{H.c}.,\eea where the $\La$ is a new scale with the mass dimension. All the couplings, $h$'s, conserve $B-L$, except that $s^\nu$ violates $L$ by two unit. It can generate the small masses for the neutrinos \cite{DNS}.

Let us introduce a $Z_2$ symmetry  and all fields of the simple 3-3-1 model
are assigned even under the $Z_2$. Below, we
consider  replication of the simple 3-3-1 model by adding an extra scalar triplet, either
$\eta'$ or $\chi'$
assigned as an odd field under the $Z_2$.

\subsection{\label{inerteta}The simple 3-3-1 model with $\eta$ replication}

An extra scalar triplet that replicates  $\eta$ is defined as
 \bea \eta' &=&  \left(\begin{array}{c}
\fr{1}{\sqrt{2}}(H'_1+iA'_1)\\
\eta'^-_2\\
\eta'^{+}_3\end{array}\right)\sim (1,3,0).\label{vev3}
\eea We notice that the $\eta'$ and $\eta$ have the same gauge quantum numbers. However,
$\eta'$ is assigned as an odd field under the $Z_2$, $\eta'\rightarrow - \eta'$, so
$<\eta'>=0$.

The scalar potential includes the $V_{\mathrm{simple}}$ given in Eq. (\ref{Vsimple})
and the terms contained $\eta'$,   \bea
V_{\eta'}&=&\mu^2_{\eta'}
\eta'^\dagger \eta' +  x_1 (\eta'^\dagger \eta')^2+x_2(\eta^\dagger
\eta)(\eta'^\dagger \eta')+x_3(\chi^\dagger \chi)(\eta'^\dagger
\eta')\crn &&+x_4(\eta^\dagger \eta')(\eta'^\dagger \eta)
+x_5(\chi^\dagger \eta')(\eta'^\dagger\chi) +\fr 1 2
[x_{6}(\eta'^\dagger \eta)^2+H.c.].\label{the1} \eea
Here, $\mu_{\eta'}$ has mass dimension, while $x_i$ $(i= 1,2,3,...,6)$
are dimensionless. All the $x_6$, $u$ and $\om$ can be considered to be real.

The model requires \cite{inert331} \be
\mu^2_{\eta'}>0,\hs x_{1,3}>0,\hs x_2+x_4\pm x_6>0.\ee
The gauge states $H'_1$, $A'_1$, $\eta'^\pm_{2} \equiv H'^\pm_{2}$ and
$\eta'^\pm_{3}\equiv H'^\pm_{3}$ by themselves are physically inert particles
with the corresponding masses as follows
\bea m^2_{H'_1} &=& M^2_{\eta'}+\fr 1 2 (x_4+x_6)u^2,\hs m^2_{A'_1}=M^2_{\eta'}+\fr 1 2 (x_4-x_6)u^2,\crn
m^2_{H'^\pm_{2}}&=&M^2_{\eta'},\hs  m^2_{H'^\pm_{3}}=M^2_{\eta'}+ \fr 1 2 x _5 \om^2,
\label{MetaP}\eea
where $M^2_{\eta'}\equiv \mu^2_{\eta'}+ \fr{1}{2} x_2 u^2 + \fr{1}{2} x_3 \om^2$.
If $H'_1$ (or $A'_1$) is the lightest inert particle (LIP), it can be the dark matter candidate.

All the interactions in (\ref{the1}) conserve $B-L$ except the $x_6$ one, since in principle the $\eta'$ fields can have arbitrary $B-L$ charges. This is analogous to the case of the 3-3-1 model with right-handed neutrinos \cite{inert331}. The masses of $H'_1$ and $A'_1$ are separated by $x_6$. Otherwise, the conservation of $B-L$, i.e. $x_6=0$, rules out the candidates $H'_1$ and $A'_1$ because they possess a large scattering cross-section off nuclei due to the $t$-channel exchange by $Z$ boson~\cite{tanxabyz}.

Let us calculate the interactions of the inert particles with the normal ones.
Due to the $Z_2$ symmetry, the inert scalars
interact only with normal scalars and gauge bosons, not with
fermions. Details of interactions are given in Appendix \ref{ttdeta}.

Under the Standard Model symmetry, the candidates $H'_1$, $A'_1$ transform as a $SU(2)_L$ doublet, which are analogous to the ones of the inert doublet model \cite{idmma}. However, our candidates are distinguishable due to the following two points: (i) Since $\om$ is the 3-3-1 breaking scale fixed at TeV range \cite{DNS}, the candidates which have masses $\sim \om$ are naturally heavy. However, also note that their masses depend on the scalar couplings as well as $\mu_{\eta'}$-parameter. (ii) Besides the interactions with the Standard Model particles, the candidates have new interactions with the new gauge and Higgs bosons.
That is to be said, in the large mass region the dark matter observables can be governed by new physics of the 3-3-1 model.

\subsection{\label{inertchi}The simple 3-3-1 model with $\chi$ replication}

The $\chi$ replication takes the form
\bea \chi' &=&  \left(\begin{array}{c}
\chi'^-_1\\
\chi'^{--}_2\\
\fr{1}{\sqrt{2}}(H'_3+iA'_3)\end{array}\right)\sim (1, 3,-1).\label{vev11}\eea
The $\chi'$ is assigned odd under the $Z_2$ symmetry that requires $<\chi'>=0$.
The additional potential into Eq. (\ref{Vsimple}) due to the $\chi'$ field is given as
 \bea
V_{\chi'}&=& \mu^2_{\chi'} \chi'^\dagger \chi' + y_1 (\chi'^\dagger \chi')^2 +y_2(\eta^\dagger \eta)(\chi'^\dagger
\chi')+y_3(\chi^\dagger \chi)(\chi'^\dagger \chi')\crn
&&+y_4 (\eta^\dagger \chi')(\chi'^\dagger\eta)+y_5(\chi^\dagger
\chi')(\chi'^\dagger \chi) +\fr 1 2 [y_{6}(\chi'^\dagger
\chi)^2+\mathrm{H.c}.]. \eea
To make sure
the scalar potential is bounded from below and the $Z_2$ is conserved by the vacuum, we impose \be \mu^2_{\chi'}>0,\hs  y_{1,2}>0,\hs y_3+y_5\pm y_6>0.\ee

The physical inert scalars $H'_3$, $A'_3$, $\chi'^\pm_1 \equiv H'^\pm_1$ and $\chi'^{\pm\pm}_2\equiv H'^\pm_2$ by themselves with the respective masses are obtained as
\bea m^2_{H'_3}&=&M^2_{\chi'}+\fr 1 2 (y_5+y_6)\om^2,\hs m^2_{A'_3}=M^2_{\chi'}+\fr 1 2 (y_5-y_6)\om^2,\crn
m^2_{H'^{\pm\pm}_2}&=&M^2_{\chi'},\hs  m^2_{H'^\pm_1}=M^2_{\chi'}+\fr 1 2 y_4 u^2,  \eea
where $M^2_{\chi'}\equiv \mu^2_{\chi'}+\fr 1 2 y_2 u^2+\fr 1 2 y_3 \om^2$.
If $H'_3$ (or $A'_3$) is the  LIP, it can be the dark matter candidate.

The couplings $y_{1,2,3,4,5}$ conserve $B-L$, whereas $y_6$ violates this charge since $\chi'$ can have arbitrary $B-L$ charges. The masses of $H'_3$ and $A'_3$ are separated by $y_6$, which is similar to the previous case. If their masses are degenerate, i.e. $B-L$ is conserved, there is a scattering of $H'_3$ and $A'_3$ off nuclei due to the $t$-channel exchange by $Z'$ boson. This cross-section is also large because the $Z'$ mass is limited by the Landau pole, which is experimentally unacceptable (this matter is analogous to the case of the sextet presented in \cite{DNS}).

Let us consider the
interactions of the inert Higgs with the normal ones as well as the gauge bosons.
We remind that  the inert scalars  do not interact with fermions because of the invariance under the $Z_2$ symmetry.
Details of interactions are given in Appendix \ref{ttdchi}.

The candidates $H'_3$ and $A'_3$ transform as singlets under the Standard Model symmetry, which are similar to the phantom of Silveira-Zee model \cite{zeemo}. However, their physics is discriminated due to the interactions with the new gauge and Higgs bosons, besides the Standard Model Higgs portal. The dark matter observables in their large mass range can be governed by the new physics. Since the candidates have masses proportional to $\om$, they have natural masses in TeV range.
Please note that their masses depend on the scalar couplings as well as $\mu_{\chi'}$ parameter.

\section{\label{darkmatter}Dark matter in minimal 3-3-1 models}

Let us recall that the simple 3-3-1 model with inert $\rho$ triplet, and the model
with inert scalar sextets were previously considered in \cite{DNS}. In this work,
we study dark matter in the simple 3-3-1 model with  $\eta$
replication (called $\eta'-$model for shortcut) and the model with $\chi$
replication (called $\chi'-$model) in details.

In order to calculate the
relic density as well as indirect and direct searches for dark matter,
we use micrOMEGAs \cite{microIndir, microDir} after expanding the relevant
interactions and
implementing new model files into CalcHEP \cite{CalcHEP}.
All possible annihilation and coannihilation channels are considered in the
computation of the relic density. The coannihilation may reduce the
relic density significantly if the mass of the inert particles
exist within around 10 $\%$ or even 20 $\%$ of the LIP (lightest
inert particle) mass \cite{coanni}.

Dark matter annihilation produces pairs of the Standard Model particles (or new particles in our model) that hadronize and decay into stable particles.
Indirect search observes the signals of positrons, anti-protons, gamma-rays that are finally produced in dark matter annihilation processes. MicrOMEGAs computes
the photon, positron, anti-proton flux at a given energy $E$ and the angle in the direction of observation, which can be the source for experiments PAMELA, Fermi, and $etc$.

In direct searches, one measures the recoil energy deposited by
scattering of LIPs with the nuclei. In this work, both $\eta'-$model and $\chi'-$model provide Higgs dark matter that
can only contribute to the spin independent interaction with nuclei.
To derive the LIP-nucleus cross section we use the method, as mentioned in
\cite{microDir}. All interactions of the LIP with quarks are input
in the model files,  CalcHEP then
generates and calculates all diagrams for LIP - quark/anti-quark elastic
scattering at zero momentum. The normalized cross section on a
point-like nucleus is obtained as
\be \sigma^{SI}_{LIP-N}=\frac{4\mu^2_{LIP}}{\pi}
(Z\la_p+(A-Z)\la_n)^2, \ee
where  $\mu_{\mathrm{LIP}}$ is the LIP-nucleus reduced mass, $\mu_{\mathrm{LIP}}=m_{\mathrm{LIP}} m_{\mathrm{nuclei}}/(m_{\mathrm{LIP}} +m_{\mathrm{nuclei}})\simeq m_{\mathrm{nuclei}}$. $\la_p$ and $\la_n$ are
the   effective couplings of the LIP to protons and neutrons, respectively.
The couplings $\la_{p,n}$   are connected to the coefficients $f^N_q$, which are linked to the pion-nucleon sigma term $\sigma_{\pi N}$
and the quantity $\sigma_0$ \cite{microDir}. Recent analyses suggest that
\cite{PiN}
\be \sigma_{\pi N} = 55 - 73 \mathrm{MeV},\hs \sigma_0 = 35 \pm 5 \mathrm{MeV}.\ee
The direct rate does not change so much in the above ranges of $\sigma_{\pi N}$ and $\sigma_0$.
The results on the relic density as well as searches for dark matter in each model
are presented in subsections below.

\subsection{\label{DMeta}Dark matter in the simple 3-3-1 model with $\eta$ replication}

The inert particles in the simple 3-3-1 model with $\eta$ replication are
$H'_1, A'_1, H^{'\pm}_2, H^{'\pm}_3$. With the condition
$x_6<\mathrm{Min}\{0,\ -x_4,\ (w/u)^2x_5-x_4\}$,
$H'_1$ is the LIP and it can be a candidate for dark matter. See Appendix \ref{feynmandia} for possible (co)annihilation channels of $H'_1$.

The $\eta'-$model
contains the following parameters: $\mu^2_{\eta'},\om,\la_{1,2,3,4}, x_{1,2,3,4,5,6} $.
 Let us choose
some fixed ones as \bea &&\la_2=\la_3=\la_4= 0.1,\
x_1 = 0.01,\ x_2= 0.03,\crn && x_3 = 0.01,\ x_4 = 0.07,\ x_5 = 0.08,\ x_6 = -0.09. \label{paraEta}\eea
The coupling $\la_1$ is constrained by the mass of the Standard Model Higgs, $m_h=125$ GeV. The squared-mass splittings of the inert fields are obviously defined, where the doublet components ($H'_1, A'_1, H'_2$) are slightly separated due to the weak scale, but they are largely separated from the singlet $H'_3$ by the $\om$ scale.

From Eq. (\ref{MetaP}) the dark matter mass depends on the two parameters $\mu_{\eta'}$ and $\om$. By our choice, the $\om$ term of the dark matter mass is $\sqrt{\fr{x_3}{2}}\om\simeq 0.07\om$, which is given at the weak scale for $\om$ in a few TeV. Therefore, the dark matter mass ranges from the weak scale to TeV scale for $\mu_{\eta'}$ correspondingly varying on such range. This selection of the dark matter mass region will scan all contributions of the Standard Model and 3-3-1 ones to the dark matter relic density [(co)annihilation precesses open when the dark matter heavier than its product].

The simple 3-3-1 model inherits two distinct regions of mass spectrum: (1) given at the weak scale ($u$) of the Standard Model particles such as $t$, $h$, $Z$, $W$ and so on; (2) achieved at the TeV scale ($w$) of new particles, including $X$, $Y$, $Z'$, $J_{1,2,3}$, $H^0$, $H^\pm$. Notice that for $\om=3-5$ TeV \cite{DNS},  $X,\ Y,\ Z'$, $H^0$ (and assumed $J_{1,2,3}$) all have the mass beyond 1 TeV. But, $H^\pm$ is slightly lighter, $m_{H^\pm}\simeq 0.67-1.12$ TeV. This is due to the particular choice of the scalar couplings. Of course, one can investigate the case with all the new scalars heavy. Indeed, the conclusions given below remains unchanged.

\begin{figure}[tb]
\begin{center}
\includegraphics[width=8cm,height=6cm]{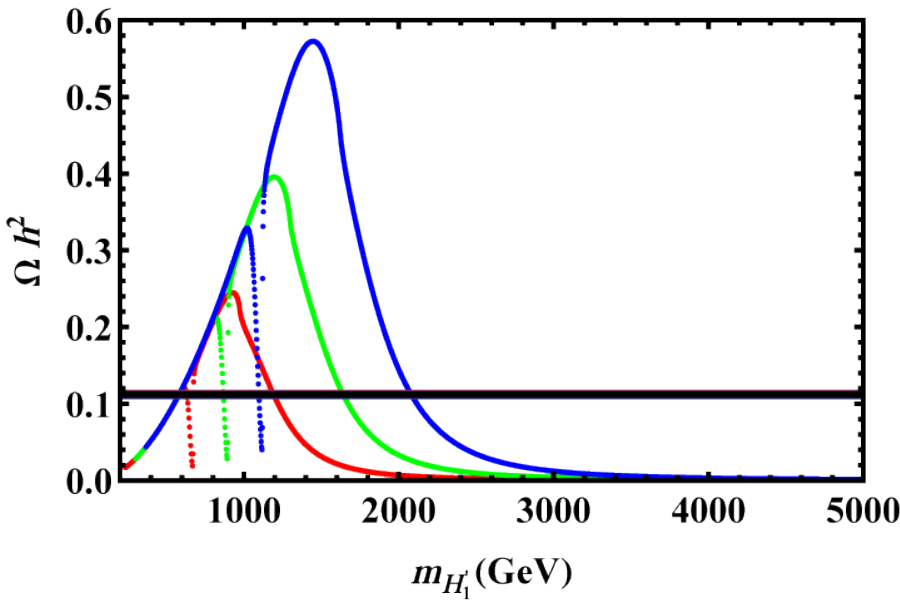}
\caption[]{\label{frelic-Eta} $\Omega h^2$ as a function of $m_{H'_1}$ for  $\om= 3$  TeV (red), $ \om= 4$  TeV (green), and $ \om= 5$  TeV (blue). (The three curved lines are coincident at low mass region and
separated at TeV scale  for $\om= 3$  TeV, $ \om= 4$  TeV, and $ \om= 5$  TeV, respectively from left to right.
The dotted lines are rare regions for $\om= 3$  TeV (left), $ \om= 4$  TeV (middle), and $ \om= 5$  TeV (right).
The horizontal line is the WMAP limit on the relic density.)}
\end{center}
\end{figure}

Fig. \ref{frelic-Eta} shows the relic density as a function of dark matter mass by varying $\mu_{\eta'}$ from 100 GeV to 5000 GeV for $\om= 3$  TeV (red), $ \om= 4$  TeV (green), and $ \om= 5$  TeV (blue). For each value of  $\om$, there are three regions of dark matter mass yielding right abundances ($\Omega h^2 \leq 0.1120 \pm 0.0056$ \cite{WMAPconstraint}, where $h$ is the reduced Hubble constant which should not be confused with the Higgs field as given at outset). \ben
\item The first region: $m_{H'_1}<600$ GeV. The relic density in this regime is governed by the Standard Model gauge and Higgs portals with only the Standard Model productions. Therefore, the relic density is independent of $\om$, the 3-3-1 breaking scale. All these can be theoretically computed which yields the effective, thermally-averaged annihilation cross-section times velocity as \cite{strumiar}
\be \langle \sigma v \rangle \simeq \left(\fr{\al}{150\ \mathrm{GeV}}\right)^2\left[\left(\fr{600\ \mathrm{GeV}}{m_{H'_1}}\right)^2+\left(\fr{x\times 1.354\ \mathrm{TeV}}{m_{H'_1}}\right)^2\right],\ee where $(\al/150\ \mathrm{GeV})^2\simeq 1\ \mathrm{pb}$, $x\equiv \sqrt{x^2_2+x^2_4+x^2_6}$, and in the brackets the first and second terms come from the gauge and Higgs portals, respectively. Because the Higgs couplings are small, $x\simeq 0.11$, the Higgs portal negligibly contributes. Hence, the relic density is governed by the gauge portal, which leads to $m_{H'_1}\simeq 600$ GeV in order to recover the correct abundance $\Omega h^2\simeq 0.1\ \mathrm{pb}/\langle \sigma v \rangle \simeq 0.11$. This matches the result given by micrOMEGAs. From Fig. \ref{frelic-Eta}, we see that the three lines coincide at the region below 600 GeV for $\om= 3$ TeV or 4 TeV or 5 TeV, as predicted. This infers that we can have a dark matter candidate with the mass just or below 600 GeV, in agreement with the WMAP results on the relic density; and it is independent with the new physics of the simple 3-3-1 model. The simple 3-3-1 model as well as the inert fields play the new role in the next two regions.

\item The second region: $H$ resonance. This regime for the dark matter relic density is very narrow, as we can be seen from  Fig. \ref{frelic-Eta} with dotted lines. It is due to a $H$ resonance through the $s$-channel annihilation of the dark matter into the Standard Model particles, including $H^\pm$ if kinematically allowed, by $H$ exchange [note that $H$ is a new Higgs of the simple 3-3-1 model]. In other words, the relic density for this regime is set by the $H$ resonance with the dark matter mass around $m_{H'_1}=\fr 1 2 m_{H}=\sqrt{\fr{\la_2}{2}}\om$, which yields $m_{H'_1}\simeq 670$ GeV for $\om=3$ TeV, $m_{H'_1}\simeq 895$ GeV for $\om=4$ TeV, and $m_{H'_1}\simeq 1.118$ TeV for $\om=5$ TeV. The resonant points (dark matter mass) as seen from the figure coincide with the given estimation. On the other hand, all the new particles of the simple 3-3-1 model are heavier than 1 TeV, except $H^\pm$ that has a mass from 670 GeV to 1.12 TeV for $\om=3-5$ TeV, aforementioned. Therefore, only the $H^\pm$ channel can be additionally opened that gives a small contribution to the relic density in this range [from 600 GeV to the point (depending on $\om$ size) before the other new particles of the simple 3-3-1 model enter the product of dark matter annihilation]. Despite of this contribution, out of the resonance regime the relic density radically increases, and overpopulates, since the dark matter mass increases.

\item The third region: 3-3-1 region. When dark matter mass reaches various masses of the new particles of the simple 3-3-1 model, the corresponding annihilation channels open, the dark matter candidate can annihilate into the new gauge bosons, new Higgs bosons, and exotic quarks. Due to the numerous contributions, the relic density decreases. It goes down to the right abundance with the values of dark matter mass evaluated as follows:
$m_{H'_1}\geq 1.15$ TeV for $\om= 3$  TeV,\ $m_{H'_1} \geq 1.6$ TeV for $\om= 4$  TeV, and $m_{H'_1} \geq 2.05$ TeV for $\om= 5$  TeV, in order to satisfy the WMAP bounds \cite{WMAPconstraint}. Reaching far above $\om$ scale, the inert fields are highly degenerated, and the coannihilations such as $H'_1 A'_1$, $H'_1 H'_2$, $H'_1 H'_3$ and so on dominate over the effective annihilation cross-section of dark matter. As a matter of fact, all the inert doublet and singlet components have gauge interactions with the ordinary and new gauge bosons (also valid for the scalar interactions, but not signified by the case) such that the $s$-channel coannihinlation cross-sections are turned on in this regime, which are more enhanced than the annihilation ones. This effect makes the relic density continuously decreasing \cite{coanni}. The simple 3-3-1 model, like the minimal 3-3-1 model, encounters a low Landau pole \cite{landaupole331}, so the next evolution of dark matter mass is nonsense.
\een

\begin{figure}[tb]
\begin{center}
\includegraphics[width=7cm,height=6cm]{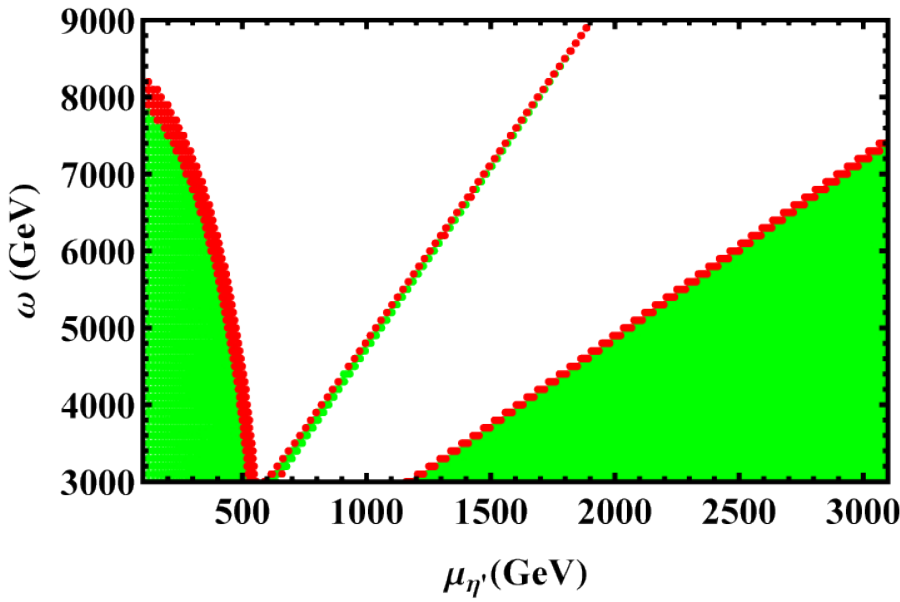}
\includegraphics[width=7cm,height=6cm]{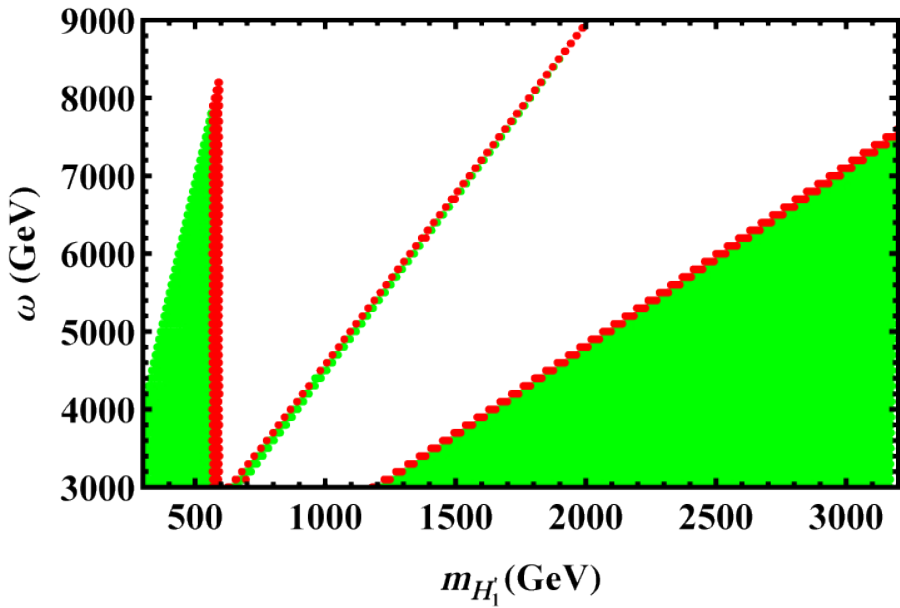}
\caption[]{\label{fwmu-Eta} Contour plot of the relic density on $\om-\mu_{\eta'}$ plane (left) and $\om-m_{H'_1}$ plane (right) in agreement with the WMAP data. The red  regions (darker fringe in black-white print) yield the right abundance, $0.1064<\Omega h^2<0.1176$.}
\end{center}
\end{figure}

All the above conclusions are more clearly shown in Fig. \ref{fwmu-Eta}, in which we figure out the plane of $\om - \mu_{\eta'}$ (left) and $\om - m_{H'_1}$
(right)
by varying both $\om$ and $\mu_{\eta'}$ in the regions (3000 GeV$<\om< $ 9000 TeV)
and (100 GeV$<\mu_{\eta'}<$ 3100 GeV). The color regions are in agreement
with the requirement $\Omega h^2<0.1176$. The red  regions satisfy
$0.1064<\Omega h^2<0.1176$. The lightest dark matter mass
can be at electroweak scale, $m_{H'_1} (min)=235.2$ GeV for $\om=3$ TeV and $\mu_{\eta'}=100$ GeV. However, please note that
this is by our choice of the parameter values despite the fact that the dark matter has a natural mass in $\om$ scale, as mentioned before.
The right panel of Fig. \ref{fwmu-Eta} shows that the 600 GeV-dark matter supplies the right relic density when $\om$ changes (corresponding to the red point-line in the region $\mu_{\eta'}<600$ GeV on the left panel). This is due to the Standard Model contribution only. The middle (straight) red point-line is due to the $H$ resonance. The rightmost red point-line is due to the contribution of new particles of the simple 3-3-1 model. Here, the dark matter mass is beyond 1 TeV.

Now let us consider the results on indirect and direct search for dark matter
in detail.
For  example, with $\om=3 $ TeV, $\mu_{\eta'}=534$ GeV we get $m_{H'_1}=574.7$ GeV and other inert particles
 $m_{H'_2}=575.2$ GeV,  $m_{A'_1}= 579.4$ GeV,  $m_{H'_3}=831.2 $ GeV.  Since the mass square difference
between $m_{H'_1}^2, m_{A'_1}^2, m_{H'_2}^2$ is in order of $x_4 u^2, x_6 u^2$, the mass of
 $A'_1$ and $ H_2^{'\pm}$ is very close to $m_{H'_1}$ for any values of $\om$. That is why the co-annihilation
contributes a lot to the $\fr 1 {\Omega h^2}$. For the choice $\om=3 $ TeV, $\mu_{\eta'}=534$ GeV, we get
$\Omega h^2= 0.111$ and the main annihilation/co-annihilation channels are
\be H'^+_2 H'^-_2 \rightarrow W^+ W^-, \ H'_1 H'_1 \rightarrow Z_1 Z_1,\
 H'_1 H'_1 \rightarrow W^+ W^-, \ H'_1 H'^\pm_2 \rightarrow A W^\pm, \  H'^+_2 H'^-_2 \rightarrow A A .\ee
In this case, the photon flux, positron flux and anti-proton flux are
$$2.8\times 10^{-14}\ (\mathrm{cm^2\ sr\ s\ GeV})^{-1},~~~
1.8 \times 10^{-12}\ (\mathrm{cm^2\ sr\ s\ GeV})^{-1},~~~
3.5 \times 10^{-11}\ (\mathrm{cm^2\ sr\ s\ GeV})^{-1},$$ correspondingly,  for the angle of sight 0.10 rad and
energy $E=100$ GeV. The $H'_1-p,n$ cross section is $1.5\times 10^{-47}\mathrm{cm^2}$
and the total number of events is $2.2\times 10^{-6}$ events/day/kg.

\begin{figure}[tb]
\begin{center}
\includegraphics[width=8cm,height=6cm]{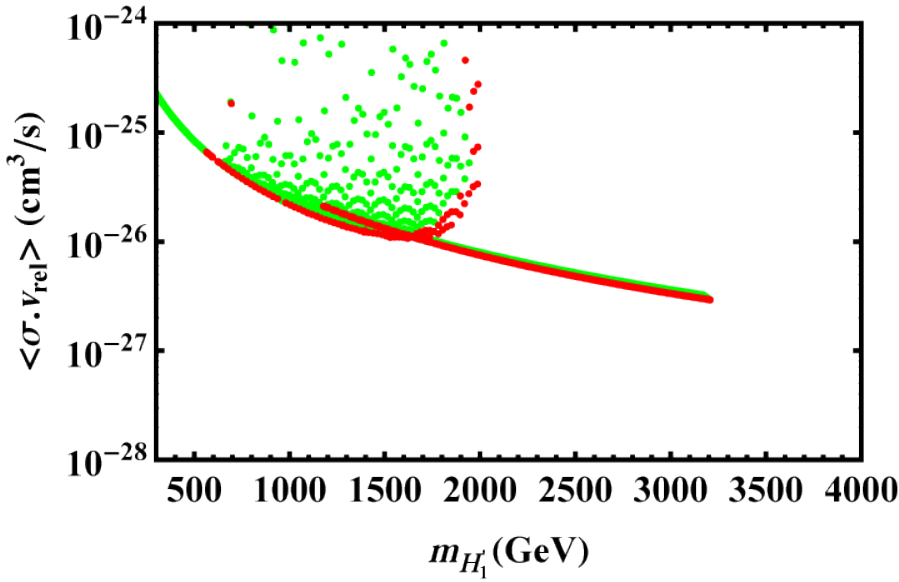}
\caption[]{\label{fsigmaV-Eta} $<\sigma .v_{rel}>-m_{H'_1}$ plane in agreement with the WMAP data.
The red regions (darker regions in black-white print) yield the right abundance, $0.1064<\Omega h^2<0.1176$.}
\end{center}
\end{figure}

The dark matter mass can be at TeV scale if we chose $\mu_{\eta'}=1171$ GeV for $\om=3 $ TeV. In this case the
dominant channels of annihilation/co-annihilation can be heavy gauge bosons, such as $H'^+_2 H'^+_3 \rightarrow W^+ X^+, \ Y^{++}Z_1$. For the dark matter with the mass around 570 GeV, the results on the relic density as well as search for dark matter do not change when varying $\om$ since the
couplings in the dominant channels do not depend on $\om$, aforementioned. The plane $<\sigma .v_{rel}>-m_{H'_1}$ for the abundance below the
experimental upper bound, $\Omega h^2 (max)=$0.1176,
is shown in
Fig. \ref{fsigmaV-Eta}.
For the right abundance of dark matter, the total annihilation cross section times the relative velocity of incoming dark matter particles and the dark matter mass  is in
order of $10^{-26} \mathrm{cm^3/s}$, respectively for $m_{H'_1} <2 $ TeV, and it
decreases when $m_{H'_1}$ increases because the heavier dark matter is, the more the contribution of
co-annihilation to the $\fr 1 {\Omega h^2}$ gets.

\begin{figure}[tb]
\begin{center}
\includegraphics[width=7cm,height=6cm]{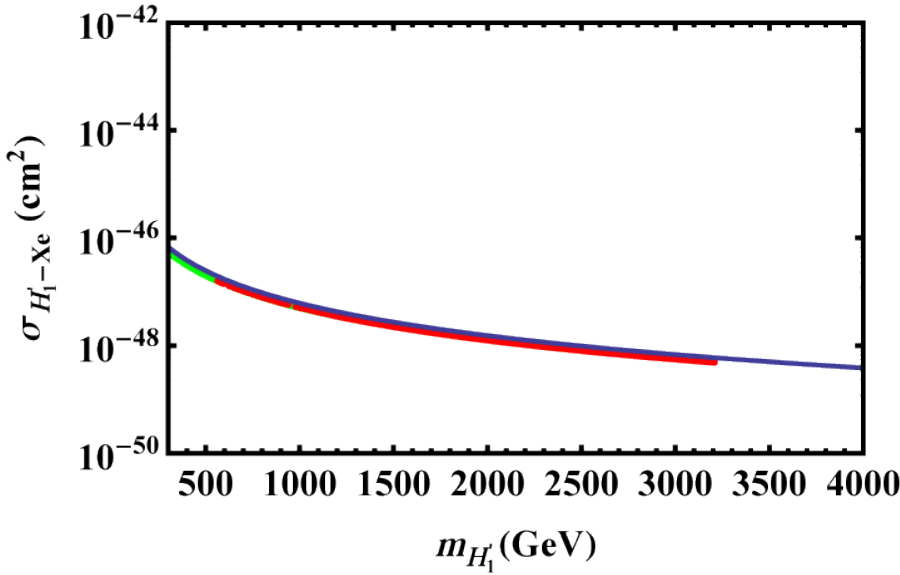}
\includegraphics[width=7cm,height=6cm]{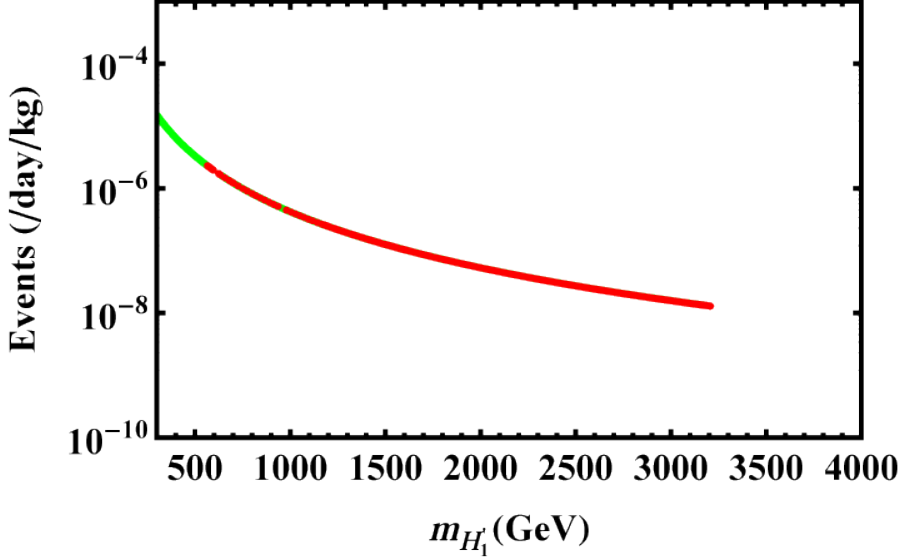}
\caption[]{\label{fsigmaN-Eta} $\sigma_{H'_1-N}$
(left) and the total number of
events/day/kg (right) as functions of $m_{H'_1}$.
The (blue) continuous line on the left panel is obtained by hand for direct search.}
\end{center}
\end{figure}

Fig. \ref{fsigmaN-Eta} shows the values of $\sigma_{\rm LIP-nucleon}$
as a function of dark matter mass obtained from micrOMEGAs by fixing the nucleon form
factors, $\sigma_0 = 30$ MeV and $\sigma_{\pi N}=73$ MeV. The value  of
$\sigma_{\rm LIP-nucleon}$ is
 $5.4\times 10^{-48} \mathrm{cm^2}$ for Xe detector and the total number of
events is $1.1\times 10^{-8}$ events/day/kg for dark matter with mass around 2 TeV.

Let us calculate the direct dark matter search by hand and compare to the results
achieved from micrOMEGAs. The dark matter scatters off the nuclei of a large detector
via interaction  with quarks
confined in nucleons. Since the dark matter is closely non-relativistic,
the process can be described by an effective Lagrangian \cite{microDir},
\be \mathcal{L_S}=2\la_q m_{H'_1} H'_1 H'_1\bar{q}q.\ee Note that,
for the real scalar field only spin-independent and even
interactions are possible. There exist interactions of the pair
$H'_1$ couple to $h$ and $H_0$. However, the dominant contributions
to $H'_1$- quark scattering are done by
the t-channel exchange of $h$.
We obtain \be \la_q=\fr{(x_2+x_4+x_6) m_q}{2m_{H'_1}
m^2_{h}}.\label{hsq}\ee
The $H'_1$-nucleon scattering amplitude is taken as a summation
over the quark level interactions with respective nucleon form
factors. The $H'_1$-nucleon cross section is given as \be
\sigma_{H'_1-N}=\fr{4m^2_{r}}{\pi}\la^2_N, \ee where $N= p,\ n$
denotes nucleon, and \bea m_r &=& \fr{m_{H'_1}
m_N}{m_{H'_1}+m_N}\simeq m_N,\crn
\fr{\la_N}{m_N}&=&\sum_{u,d,s}f^N_{Tq}\fr{\la_q}{m_q}+\fr{2}{27}f^N_{TG}\sum_{c,b,t}\fr{\la_q}{m_q}, \eea
where $f^N_{TG}=1-\sum_{u,d,s}f^N_{Tq}$. The $f^N_{Tq}$
values were considered in \cite{john},
\be f^N_{Tu}=0,014\pm
0,003, \hs f^N_{Td}=0,036\pm 0,008, \hs f^N_{Ts}=0,118\pm 0,062.\ee
Taking
$m_N=1$ GeV and $m_h=125$ GeV \cite{pdg}, we obtain \bea
\sigma_{H'_1-N} &\simeq& \left[\fr{(x_2+x_4+x_6)\ \mathrm{TeV}}{m_{H'_1}}\right]^2\times 6.146 \times 10^{-44}\
\mathrm{cm}^2 \crn &\simeq& \left[\fr{1 \
\mathrm{TeV}}{m_{H'_1}}\right]^2\times 6.146 \times 10^{-48}\
\mathrm{cm}^2,\label{sigmaH1N}\eea
with notice that the $x_{2,4,6}$ given in Eq. (\ref{paraEta}). The $\sigma_{H'_1-N}$
got in Eq. (\ref{sigmaH1N}) is inversely proportional to the square of the dark
matter mass that is shown as a (blue) continuous line passed by the red region in Fig. \ref{fsigmaN-Eta}.
It implies that the direct search calculated by hand is in nice agreement with the result yielded by micrOMEGAs
package.

Dark matter candidates can be searched at particle colliders, too. At the LHC, when the protons collide, it may produce the candidates, recognized in form of large missing transverse momentum or energy. The minimal experimental signature would be an excess of a mono-$X$ final state, recoiling against such missing. When $H'_1$ is in the first region, its production is via the exchanges of the Standard Model $h$, $Z$ and $W$ bosons as it has couplings: $hH'_1 H'_1$, $ZH'_1 A'_1$, $WH'_1 H'_2$, $hhH'_1H'_1$, $ZZH'_1 H'_1$, and $WWH'_1 H'_1$ (note that $h$ can interact with gluons via a $t$-quark loop). The mono-$X$ signatures possibly include: (i) jet, which is either a gluon $(g)$ or a quark ($q$), by processes $gg\rightarrow g H'_1H'_1$, $gq\rightarrow q H'_1 H'_1$, $q\bar{q}\rightarrow g H'_1H'_1$ (all via $h$ exchange), $gq\rightarrow q H'_1 A'_1$, $q\bar{q}\rightarrow gH'_1A'_1$ (all via $Z$ exchange), and $gq\rightarrow q H'_1 H'_2$, $q\bar{q}\rightarrow gH'_1H'_2$ (all via $W$ exchange); (ii) $Z$($W$) by process $q\bar{q}\rightarrow Z(W) H'_1 H'_1$ via $Z$($W$) and (or not) $h$ exchange; and (iii) $h$ by processes $gg\rightarrow h H'_1 H'_1$, $q\bar{q}\rightarrow h H'_1 A'_1$ via $h$ or $Z$($W$) exchange. Note that for the processes concerning $W$ boson, the two fields $(q,q)$ do not mean the same quark. When $H'_1$ is in the second or third region, the new physics of 3-3-1 model contributes, instead, where we have similar processes with $h$ replaced by $H$ and $Z$ replaced by $Z'$ (in this case, $H$ interacts with gluons via exotic quark loops). The mono-$X$ signatures are jet, $Z'$, $H$, possibly including $H^\pm$, $X^\pm$, $Y^{\pm\pm}$, and exotic quarks additionally. The LHC run I data might provide some constraints, but the LHC run II would yield crucial tests of them. All these are worth exploring to be devoted to further studies.

\subsection{\label{DMchi}Dark matter in the simple 3-3-1 model with $\chi$ replication}

The simple 3-3-1 model with $\chi$ replication contains six inert particles $H^{'\pm}_1,H^{'\pm \pm}_2,
H'_3, A'_3 $. If we assume that $y_6<\mathrm{Min}\{0,\ -y_5,\ (u/w)^2y_4-y_5\}$,
$H'_3$ is the lightest inert particle and can be the dark matter candidate. The (co)annihilation processes concerning this candidate are given in Appendix \ref{feynmandia}.

\begin{figure}[b]
\begin{center}
\includegraphics[width=8cm,height=6cm]{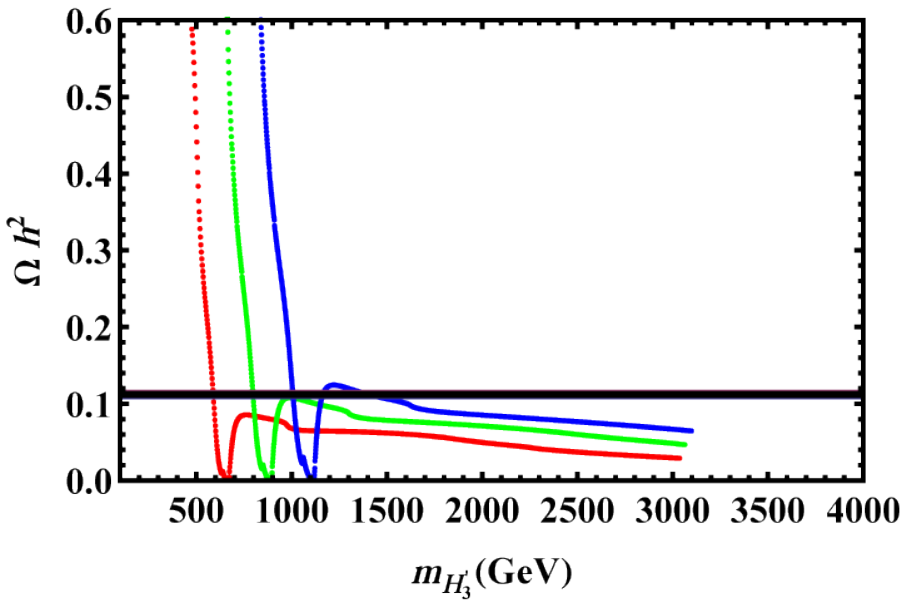}
\caption[]{\label{frelic-Chi} $\Omega h^2$ as a function of $m_{H'_3}$ for  $\om= 3$  TeV (red), $ \om= 4$  TeV (green), and $ \om= 5$  TeV (blue), respectively from left to right.}
\end{center}
\end{figure}

The parameters appeared in this model are  $\mu^2_{\chi'},\om, \la_{1,2,3,4}, y_{1,2,3,4,5,6} $,
in which the couplings $\la_{1,2,3,4}$ are fixed  as given in the $\eta'$-model.
Now let us consider the results for the relic density and indirect search as well as direct search with a set
of $y_{1,2,3,4,5,6}$ in the same order:
\be y_1 = 0.01,\ y_2= 0.04,\ y_3= 0.058,\ y_4= 0.01, y_5=0.05,\ y_6= -0.06.\label{paraChi}\ee
All the ingredients of the simple 3-3-1 model, such as the masses of new particles, the mass hierarchies among the new particles and ordinary particles, and the couplings as given, retain. Note also that  the squared-mass splittings of the inert fields are definitely small, where the doublet components $(H'_1,\ H'_2)$ are separated by $u$ scale, while the singlets $H'_3$ and $A'_3$ as well as the singlets and doublets are separated by $\om$ scale.

Because the dark matter $H'_3$ is a singlet under the Standard Model symmetry, it does not have the gauge interactions with the Standard Model gauge bosons. Therefore, at the low energy the gauge portal for dark matter (co) annihilations is suppressed. The relic density in this regime is only governed by the Higgs portal ($h$) with the Standard Model productions. The effective annihilation cross-section times velocity is obtained by \cite{inert331}
\be \langle \sigma v \rangle \simeq \left(\fr{\al}{150\ \mathrm{GeV}}\right)^2\left(\fr{y_2\times 2.2\ \mathrm{TeV}}{m_{H'_3}}\right)^2.\ee Since the chosen scalar coupling is small, $y_2=0.04$, the relic density given by $\Omega h^2\simeq 0.1\ \mathrm{pb}/\langle \sigma v \rangle \simeq 0.1\times (m_{H'_3}/88\ \mathrm{GeV})^2$ is overpopulated, which spoils the WMAP bounds, provided that $m_{H'_3}$ is larger than the weak scale. Of course, we can have a low energy solution for the dark matter candidate if the $y_2$ coupling is enhanced. Whilst this possibility is interesting as actually studied in the literature \cite{zeemo}, it will be neglected in our work. What concerned is the high energy regime of the dark matter, where the simple 3-3-1 model contributions become important.

The relic abundance is considered as a function of $m_{H'_3}$, shown in Fig. \ref{frelic-Chi}
for $\om= 3$  TeV (red), $ \om= 4$  TeV (green), and $ \om= 5$  TeV (blue). For each value of $\om$, when the dark matter mass rises from the outset, the relic density is rapidly decreased. This phenomenon is due to the $H$ resonance of the dark matter annihilation ($H'_3$) into the Standard Model particles, including $H^\pm$ if kinematically allowed, which is analogous to the previous model. That is to be  said, the $H$ resonance is crucial to determine the dark matter relic density at its low mass regime before the new particles of the simple 3-3-1 model enter the productions. The resonant point is given by $m_{H'_3}=\fr 1 2 m_H$ that yields $m_{H'_3}\simeq 670$ GeV for $\om=3$ TeV, $m_{H'_3}\simeq 895$ GeV for $\om=4$ TeV, and $m_{H'_3}\simeq 1.118$ TeV for $\om=5$ TeV (these values coincide with those of the previous model, respectively). Furthermore, the dark matter mass
is bounded by $m_{H'_3}\geq 580$ GeV for $\om= 3$  TeV or $m_{H'_3}\geq 770$ GeV for $\om= 4$  TeV or
$m_{H'_3}>$ 990 GeV for $\om=$ 5TeV.

After the resonant point, the relic density increases as the dark matter mass increases. But, it is quickly depopulated due to the new contributions of the simple 3-3-1 model. From the figure we see that there is a gap (in the dark matter mass) when $\om>4$ TeV, the relic density is overpopulated. The phenomenon similarly happens as the previous model since the dark matter mass increases against the contributions from the new particles of the simple 3-3-1 model. Going far above the $\om$ scale, the relic density still decreases. This effect is due to the large contributions of the coannhihilations resulting from strongly-degenerate inert fields \cite{coanni}. Since the model has a low Landau pole as mentioned \cite{landaupole331}, continuously rising the mass parameter is simply nonsense.

\begin{figure}[tb]
\begin{center}
\includegraphics[width=7cm,height=6cm]{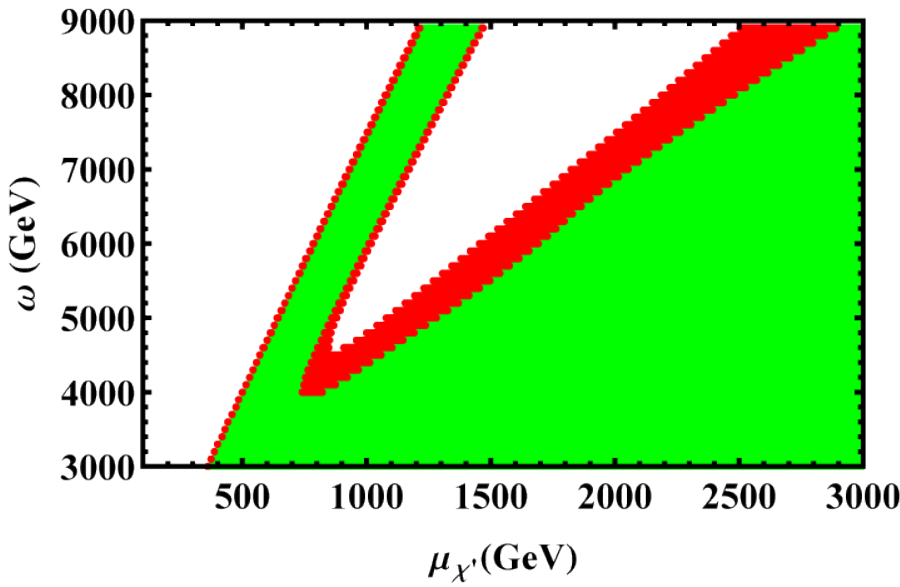}
\includegraphics[width=7cm,height=6cm]{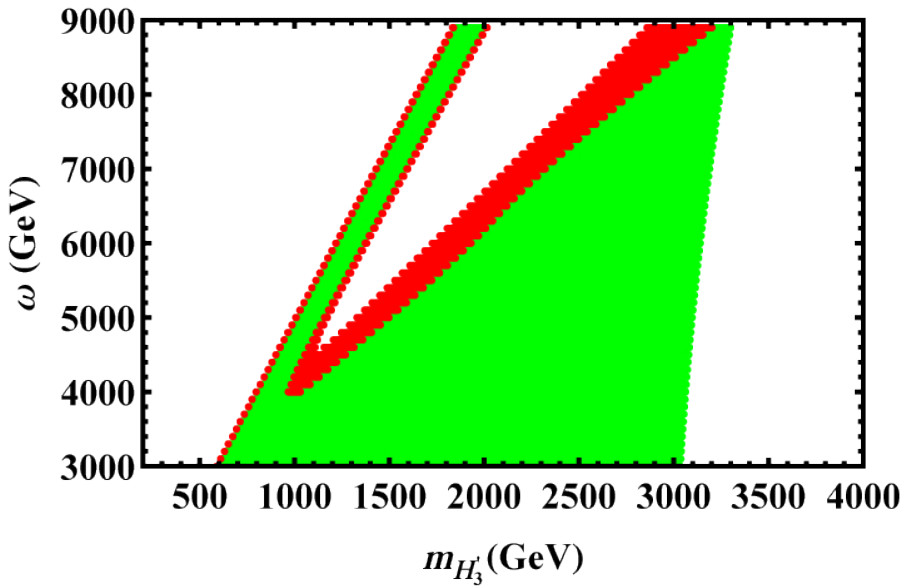}
\caption[]{\label{fwmu-Chi} Contour plot of the relic density on  $\om-\mu_{\chi'}$ plane (left) and $\om-m_{H'_3}$ plane (right) in agreement with the WMAP data. The red  regions (darker fringe in black-white print) yield the right abundance, $0.1064<\Omega h^2<0.1176$.}
\end{center}
\end{figure}

All the above discussion can be illustrated more clearly in the $\om-\mu_{\chi'}$ plane (left) and $\om-m_{H'_3}$ plane
(right) in Fig. \ref{fwmu-Chi}. For each value of $\om$, there is a lower bound on the
value of $\mu_{\chi'}$ that
results a respectively lower bound on $m_{H'_3}$ in order to satisfy the WMAP data. It is
different from the $\eta'-$model that the doublet  dark matter $H'_1$ in
the $\eta'-$model can appear near electroweak scale as governed by the Standard Model gauge portal, but the singlet one $H'_3$
in the $\chi'-$model does not happen since the gauge portal does not work. Note that in this regime both models have the suppressed Higgs portals. Given the scalar couplings are enhanced (by other choices) comparably to the gauge couplings, their dark matter phenomenologies should happen similarly. Again from the figure, the two parallel red point-lines at the leftmost present the edges of the resonant wide imposed by the WMAP bounds. The bottom of the red hat is the bound on $\om$ ($\sim$ 4 TeV) at which the relic density becomes overpopulated after the resonance. The wide red bank describes various contributions of the new particles of the simple 3-3-1 model.

\begin{figure}[tb]
\begin{center}
\includegraphics[width=8cm,height=6cm]{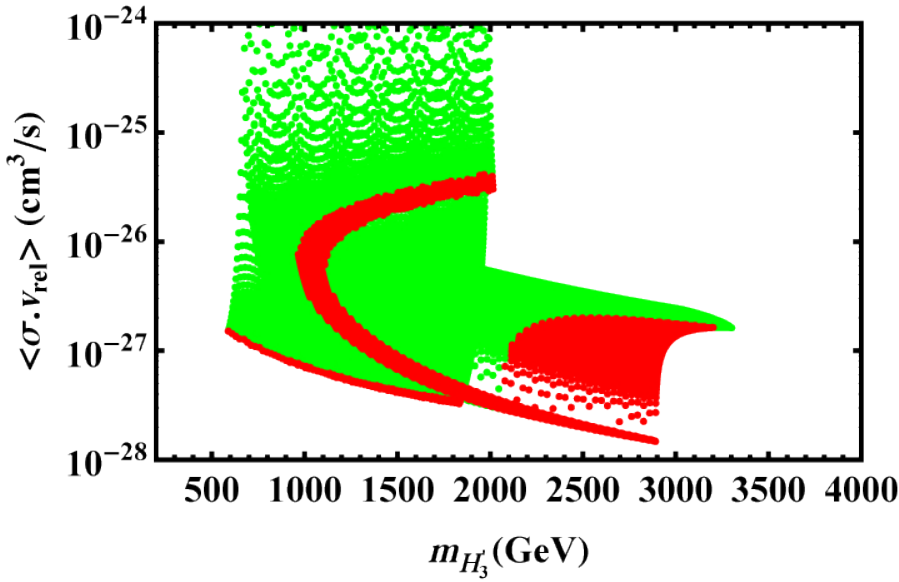}
\caption[]{\label{fsigmaV-Chi} $<\sigma .v_{rel}>-m_{H'_3}$ plane in agreement with the WMAP data.
The red regions (darker regions in black-white print) yield the right abundance, $0.1064<\Omega h^2<0.1176$.}
\end{center}
\end{figure}

\begin{figure}[tb]
\begin{center}
\includegraphics[width=7cm,height=6cm]{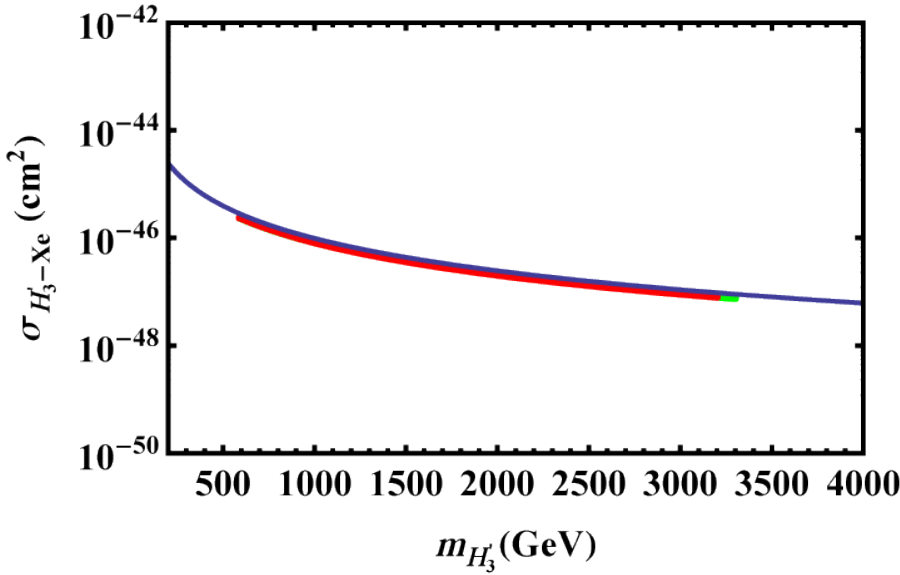}
\includegraphics[width=7cm,height=6cm]{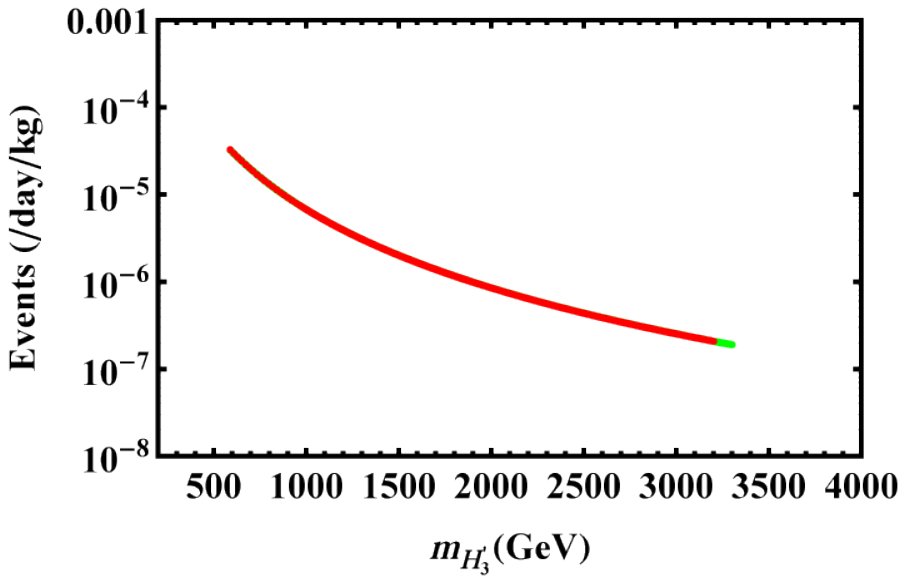}
\caption[]{\label{fsigmaN-Chi} $\sigma_{H'_3-N}$ (left) and the total number of
events/day/kg (right) as functions of $m_{H'_3}$. The (blue) continuous line on the left
panel is obtained by hand for direct search.}
\end{center}
\end{figure}

By varying $\om$ and $\mu_{\chi'}$ in the ranges (3000, 9000) GeV and  (100, 3000) GeV,
correspondingly, we figure out
the $<\sigma .v_{rel}>-m_{H'_3}$ plane in Fig. \ref{fsigmaV-Chi},
in which the
green regions satisfy the relic density $\Omega h^2 \leq0.1064$, while the red ones
yield the right abundance.
The $<\sigma .v_{rel}>$ gets the typical value $\sim 10^{-26} \mathrm{cm^3/s}$ for the
dark matter mass below 2 TeV that is similar in the $\eta'-$model.
The direct search results depending on $m_{H'_3}$ are shown in Fig. \ref{fsigmaN-Chi}.
The $H'_3-$nucleon cross section is  $2.1\times 10^{-47}\mathrm{cm^2}$ and the
number of events is $8.7\times 10^{-7}$ events/day/kg for $m_{H'_3}=2$ TeV.

Here, we give an example for the dark matter at low energy. For $\om=3$ TeV, $\mu_{\chi'}=361$ GeV,
the dark matter with mass 589 GeV  provides the abundance 0.11.
The main annihilation/co-annihilation channels are
\be
H'_3 H'_3\rightarrow  h h,\hs H'^{++}_2 H'^{--}_2 \rightarrow  h h, \hs
H'^+_1 H'^-_1 \rightarrow  h h, \hs H'_3 H'^\pm_1 \rightarrow Z_1 X^\pm, \hs
 H'_3 H'^{\pm \pm}_2 \rightarrow  Z_1 Y^{\pm\pm}.
\ee
The photon flux,  positron flux and anti-proton flux are
$$5.3\times 10^{-16}\ (\mathrm{cm^2\ sr\ s\ GeV})^{-1},~~~
2.4 \times 10^{-14}\ (\mathrm{cm^2\ sr\ s\ GeV})^{-1},~~~
6.9 \times 10^{-13}\ (\mathrm{cm^2\ sr\ s\ GeV})^{-1},$$ correspondingly, for the angle of sight 0.10 rad and
energy $E=100$ GeV. The $H'_3-p,n$ cross section is $2.3\times 10^{-46}\mathrm{cm^2}$
and the total number of events/day/kg is $3.3\times 10^{-5}$.
For the same dark matter mass around 580 GeV, the signals in indirect search for dark matter
in the $\eta'-$model are more sensitive but the direct search results are  lower than
that in the $\chi'-$model. This conclusion keeps the same if we test for
the dark matter in TeV range.

Similarly, we can calculate the direct search by hand as analysis in the $\eta'-$ model.
The effective lagrangian takes the form
\be \mathcal{L'_S}=2\la'_q m_{H'_3} H'_3 H'_3\bar{q}q.\ee
We obtain \be \la'_q=\fr{y_2 m_q}{2m_{H'_3} m^2_{h}},\ee
and finally
\be
\sigma_{H'_3-N} \simeq \left[\fr{y_2 \ \mathrm{TeV}}{m_{H'_3}}\right]^2\times 6.146 \times 10^{-44}\
\mathrm{cm}^2.\label{sigmaH3N}\ee
The result given in Eq. (\ref{sigmaH3N}) is depicted as a (blue) continuous line shown on the left side
in Fig. \ref{fsigmaN-Chi} for $y_2=0.04$. The line goes through the red region that indicates the
result calculated by hand is in nice agreement with the one yielded from micrOMEGAs.

The mono-$X$ search for $H'_3$ differs from the previous case for the low energy regime (if it is allowed by reselecting parameter values) since it has only the Standard Model Higgs portal interactions. Therefore, only the processes that are exchanged by $h$ are available. When $H'_3$ is at the high energy regime with the 3-3-1 contributions, the mono-X signatures are jet, $H$, $Z'$, or possibly include exotic quarks, $H^\pm$, $X^\pm$, $Y^{\pm\pm}$. Here, all the processes that are analogous to the previous case present. Therefore, the inert scalar singlet has rich phenomenologies featured for the 3-3-1 model, which is unlike the previous proposals. Also, the new charged scalars $H^\pm$ might present characteristic signatures at colliders since this particle is bilepton. It can be created in pair or in associated with other bileptons such as the exotic quarks and new non-Hermitian gauge bosons.

\section{\label{conclusion}Conclusion}

The minimal 3-3-1 model can work as the simple 3-3-1 model with
two scalar triplets $\eta$ and $\chi$, while leaves all other scalars as odd (inert) fields under a $Z_2$ symmetry \cite{DNS}. As a common feature of the 3-3-1 models recently investigated, the simple 3-3-1 model is only a low energy effective theory such that $B-L$ nonconserving interactions must present \cite{dongdm,inert331}. This feature is strongly supported by the fact that the proton decay operator always disappears due to the lepton-party $(-1)^L$ conservation, while the small neutrino masses result from the approximate lepton-number symmetry \cite{DNS}. Furthermore, with such criteria the inert fields as mentioned are naturally accommodated. Indeed, their presence (besides the neutrino mass operators) not only makes the model viable, but also provides dark matter candidates. The $B-L$ nonconserving interactions between the inert fields and normal scalars are crucial to determine the dark matter mass splitting from its complex counterpart. As a result, this splitting suppresses the large scattering magnitudes of dark matter with nuclei via the $Z,\ Z'$ boson exchanges (which evades the strengthen direct search bounds). Among the inert fields proposed, it is realized that the simple 3-3-1 model with inert $X=1$ sextet, the model with $\eta$ replication ($\eta'-$model) as well as the version with $\chi$ replication ($\chi'-$model) satisfy the above conditions. The last two models have been further discussed in this work.

As a matter of the fact, the original simple 3-3-1 model does not contain dark matter. The
introductory of the inert triplets ($\eta'$ in the $\eta'-$model and $\chi'$ in the $\chi'-$model)
that is odd under a $Z_2$ symmetry (all the other fields are even) makes them do
not mix with the normal ones. Due to the $Z_2$ conservation, the inert scalars have zero VEV, interact only with the normal scalars and
gauge bosons. There is no interaction between
the inert particles and fermions. The lightest and neutral inert particle is stable and
it can be the dark matter candidate. Our proposals provide a doublet dark matter $H'_1$ as in the $\eta'-$model as well as a singlet dark matter $H'_3$ as in the $\chi'-$model. All the relevant interactions, which can contribute
to the annihilation/co-annihilation processes, have been calculated. The
results for the relic density as well as experimental searches for the dark matter candidates have been
investigated by using micrOMEGAs package with the implement of the new
model files.

It is interesting that in both  $\eta'$ and $\chi'$ models, the dark matter observables in the middle scale (between the weak and 3-3-1 scales) are governed by the $H$ resonance, where $H$ is the new neutral Higgs boson of the simple 3-3-1 model. The dominant contributions from various new gauge portal and new Higgs portal set the dark matter observables in the 3-3-1 scale. The large coannihilation effects due to strongly-degenerated inert fields make the relic densities continuously decreasing when the dark matter masses are very large, far above $\om$ scale. There is a limit for the dark matter mass as well as the dark matter observables due to the Landau pole subjected to the 3-3-1 models. At the low energy, the doublet candidate $H'_1$ has the Standard Model gauge portal interactions, whereas the singlet one $H'_3$ does not. Both candidates can interact with the Standard Model via the Higgs portal ($h$). If the scalar couplings for the candidates are small in comparable to the gauge couplings, there is no low energy solution for the singlet candidate. However, when the scalar couplings become comparable to the gauge ones, all of them can be realized as low energy dark matters via the Higgs portal language.

Upon the parameter values imposed, the following conclusions are derived:

\ben
\item The region below 2 TeV yields the typical value of the thermally-averaged annihilation cross section times velocity for dark matter, $<\sigma v>\sim 10^{-26}\mathrm{cm^3/s}$.
\item The dark matter-nucleon scattering cross-section given by micrOMEGAs perfectly coincides with
the theoretical computation. Furthermore, the values achieved is in agreement with the experimental data.
\item For each value of $\om$, the dark matter mass region that yields the correct
abundance is quite narrow.
\item For all values of $\om$, the doublet dark matter $H'_1$ in the $\eta'-$model can be at the electroweak
scale, up to the one bounded by $600$ GeV. The singlet dark matter $H'_3$ in the $\chi'-$model disappears in this range. There is
a lower bound on $m_{H'_3}$, for example $m_{H'_3}>$ 580 GeV for $\om=$ 3TeV or
$m_{H'_3}>$ 770 GeV for $\om=$ 4TeV or
$m_{H'_3}>$ 990 GeV for $\om=$ 5TeV.
\item Both the models have same resonance point at the middle scale, $m_{\mathrm{DM}}=\fr 1 2 m_H$, which are $m_{\mathrm{DM}}\simeq 670$ GeV for $\om=3$ TeV, $m_{\mathrm{DM}}\simeq 895$ GeV for $\om=4$ TeV, and $m_{\mathrm{DM}}\simeq 1.118$ TeV for $\om=5$ TeV.
\item The indirect search (the particle fluxes) for the dark matter candidate in the $\eta'-$model is more sensitive. But, the direct search results such as the  $\sigma_{\rm LIP-nucleon}$, the total number of events/day/kg,
are lower for the same dark matter mass in comparison with those signals in the $\chi'-$model.
\een

With the results obtained, we conclude that the 3-3-1 models may have a natural room for dark matter, and the dark matter phenomenologies are rich. All these call for further studies.

\section*{Acknowledgments}

This research is funded by Vietnam National Foundation for Science and Technology Development (NAFOSTED)
under grant number 103.01-2013.43, and by the National Research Foundation of Korea (NRF)
grant funded by Korea government of the Ministry of Education, Science and
Technology (MEST) (No. 2011-0017430) and (No. 2011-0020333).
\\

\appendix
\section{\label{ttdeta} Interactions of the inert and normal sectors in the $\eta'$-model}

The Higgs boson-inert scalar interactions  are obtained by expanding the $V_{\eta'}$ as follows
 \bea
V_{\eta'} &\supset&x_1\left[\frac{1}{2}(H'^2_1+A'^2_1)+H'^+_{2}H'^-_{2}+
H'^{+}_{3}H'^{-}_{3}\right]^2
\crn && +x_2\left[\frac{1}{2}(u+h)^2+H^+H^-\right]\times\left[\frac{1}{2}(H'^2_1+A'^2_1)+H'^+_{2}H'^-_{2}+
H'^{+}_{3}H'^{-}_{3}\right]\crn &&+
\frac{x_3}{2}\left(\om+H\right)^2
\times\left[\frac{1}{2}(H'^2_1+A'^2_1)+H'^+_{2}H'^-_{2}+
H'^{+}_{3}H'^{-}_{3}\right]\crn &&
 +x_4\left[\frac{1}{2}(u+h)(H'_1+iA'_1)+H'^+_{3}H^-\right]\times\left[\frac{1}{2}(u+h)(H'_1-iA'_1)+H^+H'^-_3\right]\crn && +
\frac{x_5}{2}\left(\om+H\right)^2H'^+_{3}H'^-_{3}
+\frac{1}{2}x_6\left[[\frac{1}{2}(u +
h)(H'_1-iA'_1)]^2 +(H'^-_{3}H^+)^2+\mathrm{H.c}.\right]. \crn \label{scalar-inert1}
\eea

All the interactions of the inert scalars with the normal Higgs bosons are listed in  Table \ref{app1}. Note that the symmetry factor and imaginary unit as imposed  by the Feynman rules are not included  in the
tables (the interacting Lagrangian is understood as coupling times vertex, respectively).
\begin{table}
\bc \caption{\label{app1} Interactions of the inert scalars with the normal Higgs bosons in the $\eta'-$model.}
\vs
\begin{tabular}{ |c|c|c|c|  }
  \hline
  Vertex & Coupling &Vertex & Coupling  \\ \hline
 $h A'_1 A'_1 $
 &  $ \fr{(x_2 + x_4 - x_6) u}2$ & $h H'_1 H'_1 $
 &  $ \fr{(x_2 + x_4 + x_6) u}2$  \\ \hline
 $h H'^+_2 H'^-_2 $
 &  $ x_2  u$ & $h H'^+_3 H'^-_3 $
 &  $ x_2  u$  \\ \hline
 $H A'_1 A'_1 $ &  $\fr{x_3\om}2$ & $H H'_1 H'_1 $ &  $\fr{x_3\om}2$\\ \hline
 $H H'^+_2 H'^-_2 $
 &  $ x_3 \om$ & $H H'^+_3 H'^-_3 $
 &  $ (x_3 + x_5) \om$  \\ \hline
 $H'_1H^+H'^-_3$ & $\fr {(x_4 + x_6)u}2$& $A'_1H'^+_3 H^-$ & $\fr {i(x_6 - x_4)u}2$\\ \hline
 $H'_1 H'_1 h h $ & $\fr{x_2 + x_4 + x_6}4$ & $H'_1 H'_1 H^+H^- $ & $\fr{x_2}2$ \\ \hline
 $H'_1 H'_1 H H $ & $\fr{x_3}4$ & $H'_1 H'_1 A'_1 A'_1 $ & $\fr{x_1}2$\\ \hline
 $A'_1 A'_1 H^+H^- $ & $\fr{x_2}2$ & $A'_1 A'_1 H H  $ & $\fr{x_3}4$\\ \hline
 $A'_1 A'_1 h h $ & $\fr{x_2 + x_4 - x_6}4$ & $A'_1 A'_1 H'^+_2 H'^-_2  $ & $x_1$\\ \hline
 $A'_1 A'_1 H'^+_3 H'^-_3 $ & $x_1$ & $H'_1 H'^+_3 H^- h  $ & $\fr{x_4+x_6}2$\\ \hline
 $A'_1 H'^+_3 H^- h  $ & $\fr {i(x_6-x_4)}2$ & $H'_1 H'_1 H'^+_2 H'^-_2 $ & $x_1$ \\ \hline
 $h h H'^+_2 H'^-_2   $ & $\fr{x_2}2$ & $H H H'^+_2 H'^-_2   $ & $\fr{x_3}2$  \\ \hline
 $H^+H^- H'^+_2 H'^-_2   $ & $x_2$ & $H'^+_2 H'^-_2 H'^+_3 H'^-_3$ & $2x_1$ \\ \hline
 $H'_1 H'_1 H'^+_3 H'^-_3 $ & $x_1$& $h h H'^+_3 H'^-_3  $ & $\fr{x_2}2$\\ \hline
 $H H H'^+_3 H'^-_3  $ & $\fr{x_3+x_5}2$& $H^+H^- H'^+_3 H'^-_3  $ & $x_2+x_4$ \\ \hline
\end{tabular}
\ec
\end{table}

\begin{table}
\bc \caption{\label{app2}Triple interactions of the inert scalars with gauge bosons in the $\eta'-$model.}
\vs
\begin{tabular}{ |c| c|c|c| }
  \hline
  Vertex & Coupling & Vertex & Coupling\\ \hline
 $Z_{1\mu} H'_1 \overleftrightarrow{\partial}^\mu A'_1 $& $\fr g{2c_W}$&
 $Z_{2\mu} H'_1 \overleftrightarrow{\partial}^\mu A'_1 $& $\fr {g\sqrt{1-4s_W^2}}{2\sqrt3 c_W}$ \\ \hline
 $W_{\mu}^- H'_1 \overleftrightarrow{\partial}^\mu H'^+_2$ & $\fr{ig}2$ &
 $X_{\mu}^- H'_1 \overleftrightarrow{\partial}^\mu H'^+_3$ & $-\fr{ig}2$ \\ \hline
 $W_{\mu}^+ A'_1 \overleftrightarrow{\partial}^\mu H'^-_2$ & $-\fr{g}2$ &
 $X_{\mu}^+ A'_1 \overleftrightarrow{\partial}^\mu H'^-_3$ & $-\fr{g}2$ \\ \hline
 $A_{\mu} H'^+_2 \overleftrightarrow{\partial}^\mu H'^-_2$ & $igs_W$&
 $Y_{\mu}^{--} H'^+_2 \overleftrightarrow{\partial}^\mu H'^+_3$ & $-\fr{ig}{\sqrt2}$ \\ \hline
 $Z_{1\mu} H'^+_2 \overleftrightarrow{\partial}^\mu H'^-_2$ & $\fr{igc_{2W}}{2c_W}$ &
 $Z_{2\mu} H'^+_2 \overleftrightarrow{\partial}^\mu H'^-_2$ & $-\fr{ig\sqrt{1-4s_W^2}}{2\sqrt3 c_W}$ \\ \hline
 $A_{\mu} H'^+_3 \overleftrightarrow{\partial}^\mu H'^-_3$ & $igs_W$&
 $Z_{1\mu} H'^+_3 \overleftrightarrow{\partial}^\mu H'^-_3$ & $- igs_W t_W $ \\ \hline
 $Z_{2\mu} H'^+_3 \overleftrightarrow{\partial}^\mu H'^-_3$ & $-\fr{ig\sqrt{1-4s_W^2}}{\sqrt3 c_W}$ & & \\ \hline
\end{tabular}
\ec
\end{table}

 The triple
interactions of the two inert scalars with one gauge boson are given in\bea
\mathcal{L}^{\mathrm{triple}}_{\mathrm{gauge}-\eta'}&=&-ig  [\eta'^\dag(T_i A_{i\mu})\partial^\mu \eta']+\mathrm{H.c}.
\crn &=&
 -\fr{ig}{2}\left[\fr{1}{c_W}Z_{1\mu}
+\sqrt{\fr{1-3t^2_{W}}{3}}Z_{2\mu}\right]\frac{H'_1-iA'_1}{\sqrt{2}}
\overleftrightarrow{\pa}^\mu\frac{H'_1+iA'_1}{\sqrt{2}}\crn&&
-\fr{ig}{2}\left[-2s_W
A_\mu -\fr{c_{2W}}{c_W}Z_{1\mu}
+\sqrt{\fr{1-3t^2_{W}}{3}}Z_{2\mu}\right]H'^+_{2}
 \overleftrightarrow{\pa}^\mu H'^-_{2}\crn&&
 -ig\left[s_W A_\mu
-s_{W}t_{W}Z_{1\mu}
-\sqrt{\fr{1-3t^2_{W}}{3}}Z_{2\mu}\right]H'^{-}_{3}
\overleftrightarrow{\pa}^\mu H'^{+}_{3} \crn&&
-\fr{ig}{2}\left[W^{+}_\mu(H'_1-i A'_1)\overleftrightarrow{\pa}^\mu
H'^-_{2}+X^{-}_\mu (H'_1-iA'_1)\overleftrightarrow{\pa}^\mu H'^{+}_{3}+
\sqrt2 Y^{--}_\mu H'^+_{2}\overleftrightarrow{\pa}^\mu H'^{+}_{3}
\right.\crn&& \left.+\mathrm{H.c}.\right],\label{triple} \eea where we have
denoted $A \overleftrightarrow{\pa}^\mu B=A(\pa^\mu B)-(\pa^\mu
A)B$.

The quartic interactions of the two inert scalars with two gauge bosons
are given by \bea
\mathcal{L}^{\mathrm{quartic}}_{\mathrm{gauge}-\eta'}&=&g^2[\eta'^\dag(T_i A_{i\mu})^2\eta']\crn
&=&\fr {g^2}{4} \left[  W^{+\mu} W^-_{\mu} +  X^{+\mu} X^-_{\mu} +\fr1 2 \left(\fr1{ c_W}Z_{1\mu}+\fr{\sqrt{1 - 3 t_W^2}}{ \sqrt3 } Z_{2\mu}  \right)^2\right](H'^2_1+A'^2_1)\crn&&
+\fr {g^2}{4} \left[  2 W^{+\mu} W^-_{\mu}+ 2 Y^{++\mu} Y^{--}_{\mu} +\left( 2 s_W A_\mu +\fr{c_{2W} }{c_W } Z_{1 \mu}-\fr{\sqrt{1 - 3 t_W^2}}{\sqrt3} Z_{2 \mu}\right)^2\right] H'^+_2 H'^-_2\crn&&
+\fr {g^2}{4} \left[ 2X^{+\mu} X^-_{\mu}+2 Y^{++\mu} Y^{--}_{\mu}  + 4\left(s_W A_\mu  + s_W t_W Z_{1 \mu} +
     \fr{\sqrt{1 - 3 t_W^2}}{\sqrt3} Z_{2 \mu}\right)^2 \right]H'^+_3 H'^-_3\crn&&
+\fr {g^2}{4}\left[\left(\sqrt2 X^{-\mu} Y^{++}_{\mu}+2W^{+\mu}(-s_W  A_{\mu}   +  s_W t_W Z_{1\mu} + \fr{\sqrt{1 - 3 t_W^2}}{\sqrt3} Z_{2 \mu})\right)\right.\crn&& \left.\times(H'_1-iA'_1) H'^-_2+\mathrm{H.c}.\right]\crn&&
+\fr {g^2}{4}\left[\left(\sqrt2 W^{+\mu} Y^{--}_{\mu}+X^{-\mu}(2s_W  A_{\mu}   +  \fr{c_{2W}}{c_W} Z_{1\mu} - \fr{\sqrt{1 - 3 t_W^2}}{\sqrt3} Z_{2 \mu})\right)\right.\crn&& \left.\times(H'_1-iA'_1) H'^+_3+\mathrm{H.c}.\right]\crn&&
+\fr {g^2}{4}\left[\left( 2 W^{-\mu} X^{-}_{\mu}-\sqrt2 Y^{-- \mu}(\fr1{c_W}Z_{1\mu} +\fr{\sqrt{1 - 3 t_W^2}}{\sqrt3} Z_{2 \mu})\right) H'^+_2 H'^+_3+\mathrm{H.c}.\right].
\label{quartic}\eea
All the triple and quartic interactions of the   inert scalars with  gauge bosons
are presented in Table \ref{app2} and Table \ref{app3}, respectively.

\begin{table}\bc \caption{\label{app3}Quartic interactions of the inert scalars with gauge bosons in the $\eta'-$model.}
\vs
\begin{tabular}{ |c| c| c|c|}
  \hline
  Vertex & Coupling & Vertex & Coupling\\ \hline
 $H'_1 H'_1 W^+ W^- $  &  $\fr{g^2}4 $  & $H'_1 H'_1 X^+ X^- $  &  $\fr{g^2}4 $  \\ \hline
 $H'_1 H'_1 Z_1 Z_1 $  & $\fr{g^2}{8c_W^2} $ &
 $H'_1 H'_1 Z_1 Z_2 $  &  $\fr{g^2\sqrt{1-4s_W^2}}{4\sqrt3 c_W^2} $  \\ \hline
 $H'_1 H'_1 Z_2 Z_2 $  & $\fr{g^2(1-4s_W^2)}{24c_W^2}$ & $A'_1 A'_1 W^+ W^- $  &  $\fr{g^2}4 $ \\ \hline
 $A'_1 A'_1 X^+ X^- $  &  $\fr{g^2}4 $ &$A'_1 A'_1 Z_1 Z_1 $  & $\fr{g^2}{8c_W^2} $\\ \hline
 $A'_1 A'_1 Z_1 Z_2 $  &  $\fr{g^2\sqrt{1-4s_W^2}}{4\sqrt3 c_W^2} $ &
 $A'_1 A'_1 Z_2 Z_2 $  & $\fr{g^2(1-4s_W^2)}{24c_W^2}$ \\ \hline
 $H'_1 H'^+_2 A W^{-}$ & $-\fr{g^2s_W}2$ & $H'_1 H'^+_2 X^+ Y^{--}$ & $\fr{g^2}{2\sqrt2}$ \\ \hline
 $H'_1 H'^+_2 Z_1 W^{-}$ & $\fr{g^2s_W t_W}{2}$ &
 $H'_1 H'^+_2 Z_2 W^{-}$ & $\fr{g^2\sqrt{1-4s_W^2}}{2\sqrt3 c_W}$  \\ \hline
 $H'_1 H'^+_3 A X^{-}$ & $\fr{g^2s_W}2$& $H'_1 H'^+_3 W^+ Y^{--}$ & $\fr{g^2}{2\sqrt2}$ \\ \hline
 $H'_1 H'^+_3 Z_1 X^{-}$& $\fr{g^2 c_{2W}}{4c_W}$ &
 $H'_1 H'^+_3 Z_2 X^{-}$& $-\fr{g^2 \sqrt{1-4s_W^2}}{4\sqrt3 c_W}$\\ \hline
 $A'_1 H'^+_2 A W^{-}$ & $-\fr{ig^2s_W}2$ & $A'_1 H'^+_2 X^+ Y^{--}$ & $\fr{ig^2}{2\sqrt2}$\\ \hline
 $A'_1 H'^+_2 Z_1 W^{-}$ & $\fr{ig^2s_W t_W}{2}$ &
 $A'_1 H'^+_2 Z_2 W^{-}$ & $\fr{ig^2\sqrt{1-4s_W^2}}{2\sqrt3 c_W}$  \\ \hline
 $A'_1 H'^+_3 A X^{-}$ & $-\fr{ig^2s_W}2$& $A'_1 H'^+_3 W^+ Y^{--}$ & $-\fr{ig^2}{2\sqrt2}$ \\ \hline
 $A'_1 H'^+_3 Z_1 X^{-}$& $-\fr{ig^2 c_{2W}}{4c_W}$ &
 $A'_1 H'^+_3 Z_2 X^{-}$& $\fr{ig^2 \sqrt{1-4s_W^2}}{4\sqrt3 c_W}$\\ \hline
 $ H'^+_2 H'^-_2 A A$ & $ g^2 s_W^2$ & $ H'^+_2 H'^-_2 A Z_1$ & $ g^2 c_{2W}t_W $ \\ \hline
 $H'^+_2 H'^-_2 A Z_2 $ & $ -\fr{g^2t_W \sqrt{1-4s_W^2}}{\sqrt3} $ &
 $H'^+_2 H'^-_2 W^+W^-  $ & $\fr{g^2}2$ \\ \hline
 $H'^+_2 H'^-_2 Y^{++}Y^{--}  $ & $\fr{g^2}2$ &
 $H'^+_2 H'^-_2 Z_1 Z_1  $ & $\fr{g^2c_{2W}^2}{4c_W^2}$ \\ \hline
 $H'^+_2 H'^-_2 Z_1 Z_2  $ & $-\fr{g^2c_{2W}\sqrt{1-4s_W^2} }{2\sqrt3 c_W^2 }$ &
 $H'^+_2 H'^-_2 Z_2 Z_2  $ & $\fr{g^2(1-4s_W^2)}{12c_W^2}$ \\ \hline
 $ H'^+_3 H'^-_3 A A$ & $ g^2 s_W^2$ & $ H'^+_3 H'^-_3 A Z_1$ & $-2 g^2 s_W^2  t_W $ \\ \hline
 $H'^+_3 H'^-_3 A Z_2 $ & $ -\fr{2g^2t_W \sqrt{1-4s_W^2}}{\sqrt3} $ &
 $H'^+_3 H'^-_3 X^+X^-  $ & $\fr{g^2}2$  \\ \hline
 $H'^+_3 H'^-_3 Y^{++}Y^{--}  $ & $\fr{g^2}2$ &
 $H'^+_3 H'^-_3 Z_1 Z_1  $ & $\fr{g^2 s_W^4}{c_W^2}$ \\ \hline
 $H'^+_3 H'^-_3 Z_1 Z_2  $ & $\fr{2g^2 t_W^2\sqrt{1-4s_W^2} }{\sqrt3 }$ &
 $H'^+_3 H'^-_3 Z_2 Z_2  $ & $\fr{g^2(1-4s_W^2)}{3c_W^2}$\\ \hline
\end{tabular}
\ec
\end{table}

\section{\label{ttdchi} Interactions of the inert and normal sectors in the $\chi'$-model}

The Higgs boson-inert scalar interactions  are obtained as follows 
 \bea
V_{\chi'} &\supset&y_1\left[H'^{+}_1H'^{-}_1+
H'^{++}_2H'^{--}_2+\frac 1
2\left(H'^2_3+A'^2_3\right)\right]^2\crn &&+
y_2\left[\frac{\left(u+h\right)^2}{2}+H^+H^-\right]\times\left[H'^{+}_1H'^{-}_1+
H'^{++}_2H'^{--}_2+\frac{1}{2}(H'^2_3+A'^2_3)\right]\crn && +
\frac{y_3}{2}\left(\om+H\right)^2 \times\left[H'^{+}_1H'^{-}_1+
H'^{++}_2H'^{--}_2+\frac{1}{2}(H'^2_3+A'^2_3)\right]\crn &&
+\frac{y_4}{2}\left[(u+h)H'^{-}_1 +(H'_3+iA'_3)H^-\right]\times
\left[(u+h)H'^{+}_1+(H'_3-iA'_3)H^+\right]\crn &&+
\frac{1}{4}\left(\om+H\right)^2\left[(y_5+y_6)H'^2_3+(y_5-y_6)A'^2_3\right].
\label{scalar-inert2} \eea

Interactions of the two inert
scalars with one gauge boson are appeared in \bea
\mathcal{L}^{\mathrm{triple}}_{\mathrm{gauge}-\chi'}&=&
-ig  [\chi'^\dag(T_i A_{i\mu}-t B_\mu I)\partial^\mu \chi']\crn
&=& -\fr{ig}{2}\left[-2s_W
A_\mu +\fr{1+2s^2_{W}}{c_W}Z_{1\mu}
+\fr{1-9t^2_{W}}{\sqrt3\sqrt{ 1-3t^2_W})}Z_{2\mu}\right]H'^+_{1}
\overleftrightarrow{\pa}^\mu H'^-_{1} \crn&&
 -\frac{ig}{2}\left[-4s_W A_\mu
-c_W (1-3t^2_{W})Z_{1\mu}
+\fr{1-9t^2_{W}}{\sqrt3\sqrt{ 1-3t^2_W}} Z_{2\mu}\right]H'^{++}_{2}
\overleftrightarrow{\pa}^\mu H'^{--}_{2}\crn&&
 +ig\left[
\fr{1}{\sqrt3\sqrt{ 1-3t^2_W}}Z_{2\mu}\right]\frac{H'_3-iA'_3}{\sqrt{2}}
\overleftrightarrow{\pa}^\mu\frac{H'_3+iA'_3}{\sqrt{2}}\crn&&
-\fr{ig}{\sqrt{2}}\left[W^+_{\mu} H'^{+}_{1}
\overleftrightarrow{\pa }^\mu H'^{--}_{2} +
X^{+}_\mu\frac{H'_3-iA'_3}{\sqrt{2}}\overleftrightarrow{\pa}^\mu H'^{-}_{1}\right.\crn&&
\left.+Y^{++}_\mu\fr{H'_3-iA'_3}{\sqrt{2}}\overleftrightarrow{\pa}^\mu H'^{--}_{2}+\mathrm{H.c}.\right].\label{tripleChi}
\eea
The
quartic interactions of the two inert scalars with two gauge bosons are
given by \bea
\mathcal{L}^{\mathrm{quartic}}_{\mathrm{gauge}-\chi'}&=&g^2[\chi'^\dag(T_i A_{i\mu}-t B_\mu I)^2\chi']\crn
&=&\fr{g^2}{2}\left( W^{+\mu} W^-_\mu +
 X^{+\mu} X^-_\mu  \right)H'^+_1H'^-_1
+\fr{g^2}{2}\left( W^{+\mu} W^-_\mu +
 Y^{++\mu} Y^{--}_\mu  \right)H'^{++}_2 H'^{--}_2\crn&&
+\fr{g^2}{4} \left(-2 s_W A_\mu   + \fr{1 + 2 s_W^2}{c_W}   Z_{1\mu} + \fr{1 - 9 t_W^2}{
    \sqrt3 \sqrt{1 - 3 t_W^2}} Z_{2\mu}\right)^2  H'^+_1H'^-_1\crn&&
+\fr{g^2}{4} \left(4 s_W A_\mu   + c_W(1 -3t_W^2)   Z_{1\mu} - \fr{1 - 9 t_W^2}{
    \sqrt3 \sqrt{1 - 3 t_W^2}} Z_{2\mu}\right)^2  H'^{++}_2 H'^{--}_2\crn&&
+\fr{g^2}{4}\left[ X^{+\mu} X^-_\mu +
 Y^{++\mu} Y^{--}_\mu+ \fr{2}{3(1 -3t_W^2)}Z_2^\mu Z_{2\mu}\right](H'^2_3+A'^2_3)\crn&&
+\fr{g^2}{4}\left[2\left(X^{-\mu} Y^{++}_\mu + \sqrt2
   W^{+\mu} [-3s_W A_\mu  + 3 s_W t_W Z_{1\mu} + \fr{1 - 9 t_W^2}{
    \sqrt3\sqrt{1 - 3 t_W^2}}Z_{2\mu}]\right)H'^+_1H'^{--}_2 \right.\crn&&
+\left(\sqrt2W^{+\mu} Y^{--}_\mu +
   X^{-\mu} [-2s_W A_\mu  + \fr{1+2s_W^2}{c_W} Z_{1\mu} - \fr{1 + 9 t_W^2}{
    \sqrt3\sqrt{1 - 3 t_W^2}}Z_{2\mu}]\right)H'^+_1(H'_3+iA'_3) \crn&&
+\left(\sqrt2W^{-\mu} X^{-}_\mu +
   Y^{--\mu} [-4s_W A_\mu  -c_W (1 - 3 t_W^2) Z_{1\mu} - \fr{1 + 9 t_W^2}{
    \sqrt3\sqrt{1 - 3 t_W^2}}Z_{2\mu}]\right)
    \crn&&\times\left.H'^{++}_2 (H'_3+iA'_3)+\mathrm{H.c}.\right].
\label{quarticChi}\eea
The interactions of the inert scalars with the normal Higgs bosons in this model are given in Table \ref{app4},
while the gauge-inert field interactions are listed in Table \ref{app5} and Table \ref{app6}.

\begin{table}
\bc \caption{\label{app4}Interactions of the inert scalars with the normal Higgs bosons in the $\chi'-$model.}
\vs
\begin{tabular}{ |c|c|c|c|  }
  \hline
  Vertex & Coupling &Vertex & Coupling  \\ \hline
  $h H'_3 H'_3$& $\fr{y_2 u}2$ & $h A'_3 A'_3$ & $\fr{y_2 u}2$\\ \hline
  $h H'^+_1 H'^-_1$& $(y_2 + y_4)u$ & $h H'^{++}_2 H'^{--}_2$ & $y_2 u$\\ \hline
  $H H'_3 H'_3$& $\fr{(y_3+y_5+y_6) \om}2$ & $H A'_3 A'_3$ & $\fr{(y_3+y_5-y_6) \om}2$\\ \hline
  $H H'^+_1 H'^-_1$& $y_3 \om$ & $H H'^{++}_2 H'^{--}_2$ & $y_3 \om$\\ \hline
  $H'_3 H^- H'^+_1$& $\fr{y_4 u}2$ & $A'_3 H^- H'^+_1$& $\fr{iy_4 u}2$\\ \hline
  $H'_3 H'_3 hh$ & $\fr{y_2}4$ & $H'_3 H'_3 H^+ H^-$ & $\fr{y_2 + y_4}2$\\ \hline
  $H'_3 H'_3 H  H $ & $\fr{y_3 + y_5+y_6}4$& $H'_3 H'_3 A'_3 A'_3$ & $\fr{y_1}2$\\ \hline
  $H'_3 H'_3 H'^+_1 H'^-_1 $& $y_1$& $H'_3 H'_3 H'^{++}_2 H'^{--}_2 $& $y_1$\\ \hline
  $A'_3 A'_3 H^+ H^-$ & $\fr{y_2+y_4}2$& $A'_3 A'_3 H  H $ & $\fr{y_3 + y_5-y_6}4$\\ \hline
  $A'_3 A'_3 h h $ & $\fr{y_2}4$& $A'_3 A'_3 H'^+_1 H'^-_1 $& $y_1$\\ \hline
  $A'_3 A'_3 H'^{++}_2 H'^{--}_2 $& $y_1$& $H'_3 H'^+_1 H^- h $& $\fr{y_4}2$\\ \hline
  $A'_3 H'^+_1 H^- h $& $\fr{i y_4}2$& $h h H'^+_1 H'^-_1 $& $\fr{y_2+y_4}2$\\ \hline
  $H H H'^+_1 H'^-_1 $& $\fr{y_3}2$& $H^+ H^- H'^+_1 H'^-_1 $ &$y_2$\\ \hline
  $H'^+_2 H'^-_2 H'^+_1 H'^-_1 $ &$2 y_1$& $hh H'^{++}_2 H'^{--}_2  $ & $\fr{y_2}2$\\ \hline
  $H H H'^{++}_2 H'^{--}_2 $ & $\fr{y_3}2$ & $H^+ H^- H'^{++}_2 H'^{--}_2 $ &$y_2$\\ \hline
\end{tabular}
\ec
\end{table}

\begin{table}
\bc \caption{\label{app5}Triple interactions of the inert scalars with gauge bosons in the $\chi'-$model.}
\vs
\begin{tabular}{ |c| c|c|c| }
  \hline
  Vertex & Coupling & Vertex & Coupling\\ \hline
  $Z_{2\mu} H'_3 \overleftrightarrow{\partial}^\mu A'_3 $& $-\fr {gc_W}{\sqrt3 \sqrt{1-4s_W^2}}$&
  $X^+_{\mu} H'_3 \overleftrightarrow{\partial}^\mu H'^-_1 $& $-\fr {ig }2$ \\ \hline
  $Y^{++}_{\mu} H'_3 \overleftrightarrow{\partial}^\mu H'^{--}_2 $& $-\fr {ig }2$ &
  $X^+_{\mu} A'_3 \overleftrightarrow{\partial}^\mu H'^-_1 $& $-\fr {g }2$ \\ \hline
  $Y^{++}_{\mu} A'_3 \overleftrightarrow{\partial}^\mu H'^{--}_2 $& $-\fr {g }2$&
  $A_{\mu} H'^+_1 \overleftrightarrow{\partial}^\mu H'^-_1 $& $ igs_W $ \\ \hline
  $W^+_{\mu} H'^+_1 \overleftrightarrow{\partial}^\mu H'^{--}_2$& $ -\fr {ig }{\sqrt2} $&
  $Z_{1\mu} H'^+_1 \overleftrightarrow{\partial}^\mu H'^-_1 $& $ -\fr{ig(1+2s_W^2)}{2c_W}$ \\ \hline
  $Z_{2\mu} H'^+_1 \overleftrightarrow{\partial}^\mu H'^-_1 $& $ -\fr{ig(1-10s_W^2)}{2\sqrt3       c_W\sqrt{1-4s_W^2}}$& $A_{\mu} H'^+_2 \overleftrightarrow{\partial}^\mu H'^-_2 $& $ 2igs_W $ \\ \hline
  $W^-_{\mu} H'^{++}_2 \overleftrightarrow{\partial}^\mu H'^-_1$& $ -\fr {ig }{\sqrt2} $&
  $Z_{1\mu} H'^{++}_2 \overleftrightarrow{\partial}^\mu H'^{--}_2 $& $ \fr{ig (1-4s_W^2)}{2 c_W} $ \\ \hline
  $Z_{2\mu} H'^{++}_2 \overleftrightarrow{\partial}^\mu H'^{--}_2 $& $ -\fr{ig (1-10s_W^2)}{2\sqrt3       c_W\sqrt{1-4s_W^2}} $& & \\ \hline
\end{tabular}
\ec
\end{table}

\begin{table}\bc \caption{\label{app6}Quartic interactions of the inert scalars with gauge bosons in the $\chi'-$model.}
\vs
\begin{tabular}{ |c| c| c|c|}
  \hline
  Vertex & Coupling & Vertex & Coupling\\ \hline
  $H'_3 H'_3 Y^{++} Y^{--}$& $\fr{g^2}4$ & $H'_3 H'_3 X^+ X^-$& $\fr{g^2}4$ \\ \hline
  $H'_3 H'_3 Z_2 Z_2$& $\fr{g^2c_W^2}{6(1-4s_W^2)}$& $A'_3 A'_3 Y^{++} Y^{--}$& $\fr{g^2}4$ \\ \hline
  $A'_3 A'_3 X^+ X^-$& $\fr{g^2}4$& $A'_3 A'_3 Z_2 Z_2$& $\fr{g^2c_W^2}{6(1-4s_W^2)}$ \\ \hline
  $H'_3 H'^+_1 A X^-$& $-\fr{g^2s_W}2$& $H'_3 H'^+_1 W^+ Y^{--}$& $\fr{g^2}{2\sqrt2}$ \\ \hline
  $H'_3 H'^+_1 Z_1 X^-$& $\fr{g^2(1+2s_W^2)}{4c_W}$& $H'_3 H'^+_1 Z_2 X^-$& $-\fr{g^2(1+8s_W^2)}{4\sqrt3 c_W\sqrt{1-4s_W^2} }$ \\ \hline
  $H'_3 H'^{++}_2 W^- X^-$& $\fr{g^2}{2\sqrt2}$& $H'_3 H'^{++}_2 A Y^{--}$& $-g^2 s_W$ \\ \hline
  $H'_3 H'^{++}_2 Z_1 Y^{--}$& $-\fr{g^2(1-4s_W^2)}{4c_W}$&
  $H'_3 H'^{++}_2 Z_2 Y^{--}$& $-\fr{g^2(1+8s_W^2)}{4\sqrt3 c_W\sqrt{1-4s_W^2}}$ \\ \hline
  $A'_3 H'^+_1 A X^-$& $-\fr{ig^2s_W}2$& $A'_3 H'^+_1 W^+ Y^{--}$& $\fr{ig^2}{2\sqrt2}$ \\ \hline
  $A'_3 H'^+_1 Z_1 X^-$& $\fr{ig^2(1+2s_W^2)}{4c_W}$& $A'_3 H'^+_1 Z_2 X^-$& $-\fr{ig^2(1+8s_W^2)}{4\sqrt3 c_W\sqrt{1-4s_W^2} }$ \\ \hline
  $A'_3 H'^{++}_2 W^- X^-$& $\fr{ig^2}{2\sqrt2}$& $A'_3 H'^{++}_2 A Y^{--}$& $-ig^2 s_W$ \\ \hline
  $A'_3 H'^{++}_2 Z_1 Y^{--}$& $-\fr{ig^2(1-4s_W^2)}{4c_W}$&
  $A'_3 H'^{++}_2 Z_2 Y^{--}$& $-\fr{g^2(1+8s_W^2)}{4\sqrt3 c_W\sqrt{1-4s_W^2}}$ \\ \hline
  $H'^+_1 H'^{--}_2 A W^+$& $-\fr{3g^2 s_W}{\sqrt2}$& $H'^+_1 H'^{--}_2 X^- Y^{++}$& $\fr{g^2}2$ \\ \hline
  $H'^+_1 H'^{--}_2 Z_1 W^+$& $\fr{3g^2 s_W t_W}{\sqrt2}$&
  $H'^+_1 H'^{--}_2 Z_2 W^+$ & $\fr{g^2(1-10s_W^2)}{\sqrt6 c_W\sqrt{1-4s_W^2}}$ \\ \hline
  $H'^+_1 H'^-_1 A A$& $ g^2 s_W^2 $& $H'^+_1 H'^-_1 A Z_1$& $- g^2 t_W(1+2s_W^2) $ \\ \hline
  $H'^+_1 H'^-_1 A Z_2$& $-\fr{g^2t_W(1-10s_W^2)}{\sqrt3 \sqrt{1-4s_W^2}}$ &
  $H'^+_1 H'^-_1 W^+ W^-$& $ \fr{g^2}2 $\\ \hline
  $H'^+_1 H'^-_1 X^+ X^-$& $ \fr{g^2}2 $& $H'^+_1 H'^-_1 Z_1 Z_1$ &$\fr{g^2(1+2s_W^2)^2}{4c_W^2}$\\ \hline
  $H'^+_1 H'^-_1 Z_1 Z_2$ &$\fr{g^2(1+2s_W^2)(1-10s_W^2)}{2\sqrt3 c_W^2 \sqrt{1-4s_W^2}}$&
  $H'^+_1 H'^-_1 Z_2 Z_2$ &$\fr{g^2(1-10s_W^2)^2}{12 c_W^2 (1-4s_W^2)}$\\ \hline
  $H'^{++}_2 H'^{--}_2 A A$& $4 g^2 s_W^2$ &$H'^{++}_2 H'^{--}_2 A Z_1$& $2 g^2 t_W (1-4s_W^2)$\\ \hline
  $H'^{++}_2 H'^{--}_2 A Z_2$& $-\fr{2 g^2 t_W (1-10s_W^2)}{\sqrt3 \sqrt{1-4s_W^2}}$&
  $H'^{++}_2 H'^{--}_2 W^+ W^-$& $ \fr{g^2}2 $\\ \hline
  $H'^{++}_2 H'^{--}_2 Y^{++} Y^{--}$& $ \fr{g^2}2 $&
  $H'^{++}_2 H'^{--}_2 Z_1 Z_1$ &$\fr{g^2(1-4s_W^2)^2}{4c_W^2}$\\ \hline
  $H'^{++}_2 H'^{--}_2 Z_1 Z_2$ &$-\fr{g^2(1-4s_W^2)(1-10s_W^2)}{2\sqrt3 c_W^2 \sqrt{1-4s_W^2}}$&
  $H'^{++}_2 H'^{--}_2 Z_2 Z_2$& $\fr{ g^2 (1-10s_W^2)^2}{12c_W^2 (1-4s_W^2)}$ \\ \hline
\end{tabular}
\ec
\end{table}

\section{\label{feynmandia}Feynman diagrams}

\begin{figure}[tb]
\begin{center}
\includegraphics[width=14cm,height=18cm]{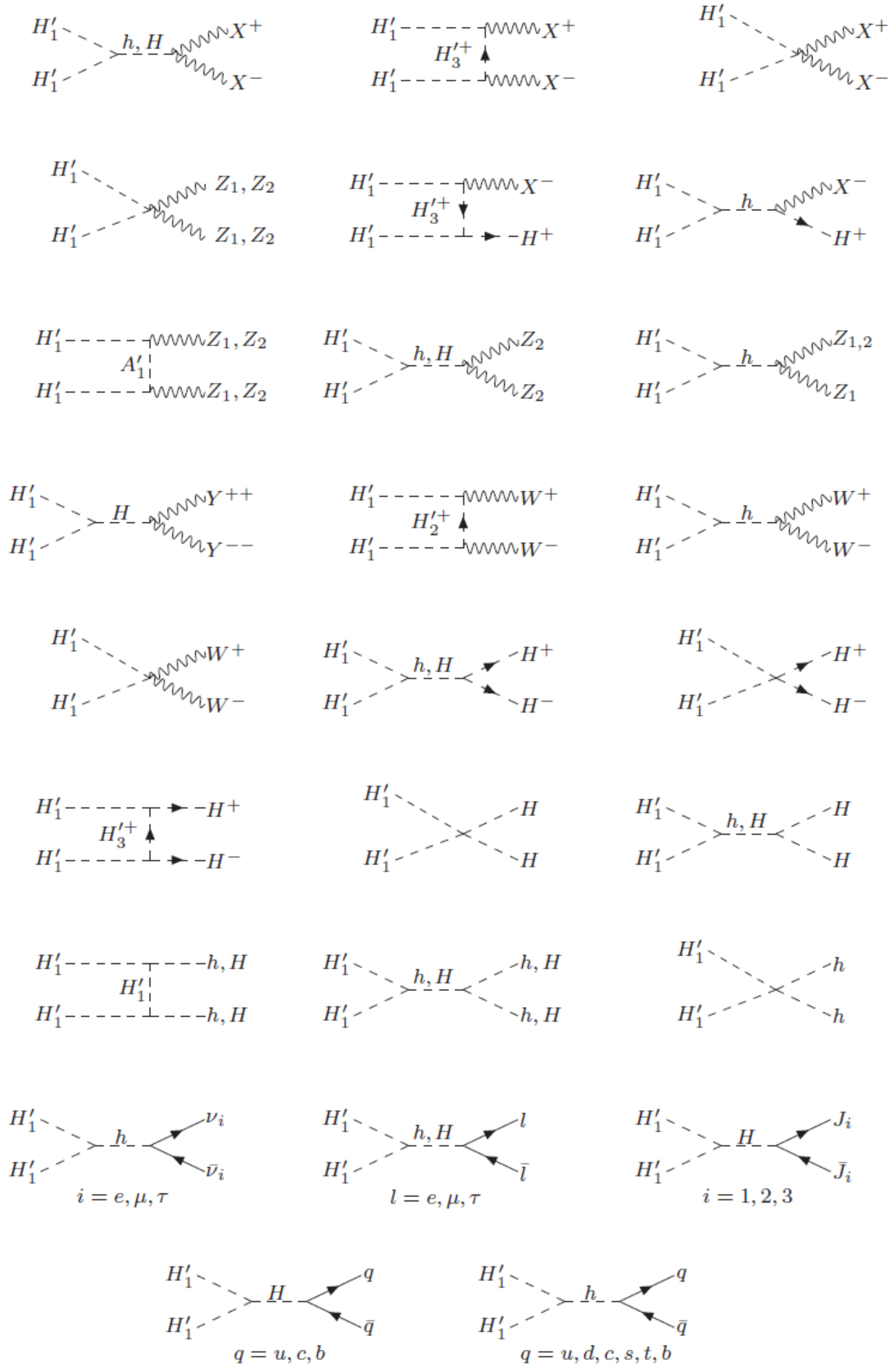}
\caption[]{\label{anniH1} Diagrams contributing to the annihilation of the $H'_1$ dark matter.}
\end{center}
\end{figure}

For a convenience in reading, we will list the Feynman diagrams for dark matter (co)annihilation processes. The annihilation channels of $H'_1$ are given in Fig. \ref{anniH1}.

\begin{figure}[tb]
\begin{center}
\includegraphics[width=14cm,height=3.5cm]{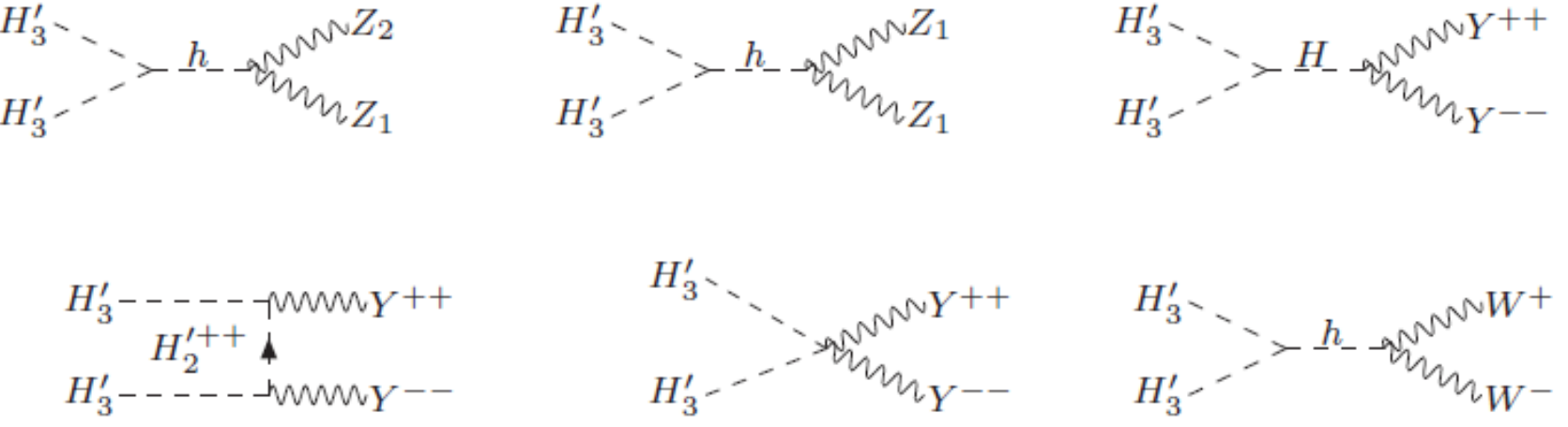}
\caption[]{\label{anniH3} Diagrams contributing to the annihilation of the $H'_3$ dark matter.
We only list the channels, which are different from ones due to
the annihilation of $H'_1$.}
\end{center}
\end{figure}

Since the candidate $H'_3$ is the Standard Model singlet, it does not interact with the Standard Model gauge bosons as $H'_1$ does. Excluding these elements, the remaining annihilation channels of $H'_1$ are almost similar to $H'_3$ by replacements: $H'_1  \rightarrow H'_3$ and $A'_1  \rightarrow A'_3$. Fig. \ref{anniH3} lists only the channels that are different from those of $H'_1$. We see that there is only one
possible diagram for each $H'_3H'_3 \rightarrow Z_1Z_1;  Z_1Z_2; W^+W^-$ via the Higgs portals, less than
the number of diagrams corresponding to $H'_1$ annihilation as commented, while there are additionally
possibilities of $H'_3H'_3 \rightarrow Y^{++}Y^{--}$.

\end{document}